\documentclass[12pt]{iopart}
\usepackage{iopams}  
\begin{document}

\title{Variational formalism for generic shells in general relativity}

\author{Bence Racskó}

\address{University of Szeged, Department of Theoretical Physics, Szeged, Hungary}
\ead{racsko@titan.physx.u-szeged.hu}
\vspace{10pt}
\begin{indented}
\item[]May 2021
\end{indented}

\begin{abstract}
We investigate the variational principle for the gravitational field
in the presence of thin shells of completely unconstrained signature
({\it generic shells}). Such variational formulations have been given
before for shells of timelike and null signatures separately, but
so far no unified treatment exists. We identify the shell equation
as the natural boundary condition associated with a broken extremal
problem along a hypersurface where the metric tensor is allowed to
be nondifferentiable. Since the second order nature of the Einstein-Hilbert
action makes the boundary value problem associated with the variational
formulation ill-defined, regularization schemes need to be introduced.
We investigate several such regularization schemes and prove their
equivalence. We show that the unified shell equation derived from
this variational procedure reproduce past results obtained via distribution
theory by Barrabes and Israel for hypersurfaces of fixed causal type
and by Mars and Senovilla for generic shells. These results are expected
to provide a useful guide to formulating thin shell equations and
junction conditions along generic hypersurfaces in modified theories
of gravity.
\end{abstract}

%
% Uncomment for keywords
\vspace{2pc}
\noindent{\it Keywords}: general relativity, thin shells, junction conditions, variational principles
%
% Uncomment for Submitted to journal title message
\submitto{\CQG}
%
% Uncomment if a separate title page is required
%\maketitle
% 
% For two-column output uncomment the next line and choose [10pt] rather than [12pt] in the \documentclass declaration
%\ioptwocol
%

\section{Introduction}

Thin shells in General Relativity (GR) and field theories in general
are weak (distributional) solutions to the field equations whose pathological
behaviour is concentrated to a single hypersurface (or a series of
nonintersecting hypersurfaces) in spacetime. In GR such solutions
may describe energetic phenomena such as phase transitions, impulsive
electromagnetic and gravitational waves \cite{Poi_Book,Bar_Hog}.
Thin shells also give rise to junction condition on the common boundary
surface when glueing together spacetime domains.\\
\\
Thin shells and junction conditions in GR have been considered by
Lanczos \cite{Lanczos}, Darmois \cite{Darmois}, O'Brien and
Synge \cite{Ob_Synge}, and Lichnerowicz \cite{Lich}, however
the most commonly used formulation has been given by Israel \cite{Israel}.
On a timelike or spacelike hypersurface partitioning the spacetime
manifold into two subdomains, Israel prescribed the continuity of
the induced metric $h_{ab}$ and the Lanczos equation relating the
jump of the extrinsic curvature $K_{ab}$ to the surface energy-momentum
tensor. In the absence of a material shell, the Lanczos equation reduces
to the continuity of the extrinsic curvature. The case when the surface
energy-momentum tensor does not vanish will be referred to as a {\it thin
shell}, and the relations imposed by the vanishing of the surface
energy-momentum tensor as the {\it junction conditions}\footnote{This terminology is not universal. Some authors refer to the Lanczos
equation itself as a junction condition, even if the surface energy-momentum
tensor does not vanish.}. An advantage of Israel's formulation is double covariance. For practical
calculations it is often useful to work with coordinate systems adapted
to the subdomains that mismatch along the hypersurface. Differentiability
classes of tensor fields may only be established in coordinate systems
whose differentiability class exceeds that of the tensor field. Israel's
equations are, however, relations between hypersurface tensors and
thus one only has to ensure that the parametrization of the hypersurface
is the same as viewed from either side and otherwise work with disjoint
systems of bulk coordinates in each spacetime region.\\
\\
Israel's formulation breaks down when the hypersurface has null points.
At null points the normal vector field becomes tangential as well
and the extrinsic curvature - which can be seen as the normal derivative
of the metric - becomes an intrinsic tangential quantity that carries
no transverse information. The 3+1 orthogonal decomposition along
the shell facilitated by the normal vector becomes degenerate. To
fix terminology, a hypersurface will be called {\it pure} if it is
either timelike, spacelike or null, while it will be referred to as
{\it causality-changing}, {\it signature-changing} or {\it non-pure} if its causal type is
not constant. The term {\it generic hypersurface} is used when the
causal type is absolutely not fixed and the surface may either be
pure or causality-changing.\\
\\
Null shells are physically relevant (we refer to \cite{Bar_Hog} for an extensive treatment of their applications), for example as models for impulsive
electromagnetic and gravitational waves. Generalizations
of the Israel formalism for null shells have been given among others
by Clarke and Dray \cite{CD}, Barrabes and Israel \cite{BI},
Mars and Senovilla \cite{MS}, Poisson \cite{Poi}, Mars \cite{Mar} and Senovilla
\cite{Sen}. Out of these, the formalisms of  \cite{BI,MS,Mar,Sen}
give a unified prescription valid for generic hypersurfaces\footnote{\label{fn:The-work-by}The work by Barrabes and Israel impose the
condition $n\cdot n=\mathrm{const}$, i.e. the length of the normal
vector is constant along the hypersurface. This formally restricts
their formalism to {\it pure} hypersurfaces. However the shell equation
obtained therein agrees (after the differences in conventions have
been addressed) by that of eg. Mars and Senovilla, which is valid
for generic shells, showing that this condition is not imperative
in the derivation of the shell equation.}. The common point of generalization is that the normal vector field
is accompanied by a transversal vector field which generates a non-orthogonal
decomposition of the spacetime along the hypersurface. The role of
the extrinsic curvature is carried by an analogous quantity built
from the transversal vector field. A hypersurface equipped with a
selected transversal vector field is called a {\it rigged hypersurface}.
This structure has been investigated by eg. Eisenhart \cite{Eisen}
and Schouten \cite{Schouten} to describe the geometry of subspaces
of manifolds with linear connections. The formalism has been systematically
applied to GR by Mars and Senovilla in  \cite{MS}.\\
\\
There exists at least four methods of obtaining the timelike or spacelike
shell equation in GR \cite{Sen}. These will be referred to as
i) the ``pillbox integration\footnote{This terminology has been borrowed from Misner, Thorne and Wheeler
\cite{MTW}.}'', ii) the distributional method, iii) the intrinsic method and iv) the variational method.
Pillbox integration has been employed by Israel in  \cite{Israel}
and involves writing the field equations in Gaussian coordinates adapted
to the hypersurface, separating a normal derivative and integrating
the field equations through the shell as its thickness tends to zero.
This approach is similar to the well-known textbook method \cite{Jack}
to derive the analogous jump conditions in electrodynamics. The distributional
method has been pioneered by Taub \cite{Taub}, Geroch and Traschen
\cite{GT}. The metric tensor is taken to be a $C^{0}$ regular
distribution\footnote{In $C^{1}$ coordinates, the continuity of the first fundamental form
is equivalent to the continuity of the spacetime metric on the shell.
See Clarke and Dray \cite{CD} as well as the comments in 
\cite{MS,Sen} for proof.}, from which it follows that the connection is allowed to have discontinuities
and the curvature tensor may contain a delta function term. The field
equations then impose a relation between the singular part of the
Einstein tensor and a singular contribution to the energy-momentum
tensor, which is Lanczos' equation. If the metric tensor were allowed
to be discontinuous, the connection would pick up a delta function
term, and the curvature tensor (quadratic in the connection) would
involve products of delta functions, which are ill-defined. This imposes
the continuity of the metric as a junction condition. The intrinsic method has been used by Mars \cite{Mar} as an application of his concept of \emph{hypersurface data}. He abstracted the properties of rigged hypersurfaces by defining data on an arbitrary hypersurface which may correspond to data specified by a rigging when the hypersurface is embedded in a pseudo-Riemannian space. The purpose has been to open the road for initial value problems in GR for any possible initial hypersurface, however through the use of the rigged analogues of the usual constraint equations, it becomes possible to formulate shells in a purely intrinsic manner with no need for even embedding the hypersurface.\\
\\
The shell equations have also been obtained via variational methods
\cite{HK,CR,Muk}. This is particularily useful for braneworld
scenarios \cite{CR}, where the Lanczos equation on the brane is
a part of the equations of motion and thus the brane and bulk dynamics
arise from a unified variational principle. When the variational formulation
is followed, the combined shell + bulk dynamics appear as the broken
extremals \cite{GH} of a variational problem with the shell equation
being the natural boundary condition on the surface.\\
\\
For a second order theory described by a first order Lagrangian, this
is straightforward. The Einstein-Hilbert Lagrangian on the other hand
is second order. Since a second order Lagrangian normally produces
equations of motion of order four, the boundary conditions pertinent
to the variational problem are that of a fourth order differential
equation and require the fixing of both the metric and its transverse
derivative at the boundary. As the actual field equations are only
second order, fixing the transverse derivative would overdetermine
the field equations and this causes the variational problem to become
ill-defined \cite{DH}. When variational principles with only outer
boundaries are considered, a common method of solution \cite{Wald}
is to add a boundary term (for example the Gibbons-Hawking-York term
\cite{York,Gib_Hawk}, but other boundary terms could be introduced
at the price of also introducing additional structures) to the Einstein-Hilbert action
such that combined bulk + boundary action requires the boundary conditions
consistent with a first order Lagrangian, and thus the variation problem
becomes well-defined. After Parattu \etal \cite{Par} such boundary
terms will be referred to as {\it variational counterterms}. A shell
may be considered as an interior boundary of the spacetime manifold
thus it is clear that some similar regularization procedure is needed
to obtain the correct results. One such way of regularization is to
also add the Gibbons-Hawking-York counterterms to the action at the
shell \cite{CR,Muk}. Another which has been employed by Hajicek
and Kijowski \cite{HK} is to consider the Lagrangian itself as
a distribution. Since the Lagrangian involves a curvature tensor,
it has a delta function term which is proportional the difference
of the Gibbons-Hawking-York terms as calculated from the two sides.\\
\\
The shell equations and junction conditions for null and generic shells
have been in general derived via the distributional method, which
is simple to generalize. Senovilla \cite{Sen} has also shown that
pillbox integration can also be adapted to the generic case, and the intrinsic method was already applicable to generic shells. It seems
however that not much attention has been given to the variational
method for hypersurfaces that are not timelike or spacelike. Jezierski,
Kijowski and Czuchry have \cite{JK} considered the variational
treatment of null shells, however they did not show that their results
agree in the null limit with the results of eg. Barrabes and Israel
\cite{BI} or Mars and Senovilla \cite{MS}. There is also an
unaddressed issue that has been pointed out by Parattu \etal \cite{Par_null}
when investigating counterterms on null boundaries. A variation in
the metric is a variation in the causality, and such variations do
not preserve the nullity of a hypersurface. The underlying reason
is that in the tangent space at a fixed point, null vectors form a
topologically closed set and every neighborhood of each null vector
contains both timelike and spacelike vectors. A general variation
will push the initially null surface off the lightcone. The same issue
is not encountered in regards to timelike or spacelike surfaces as
timelike/spacelike vectors form open sets and each such vector has
a neighborhood that consists entirely of timelike/spacelike vectors.
It stands to reason that variational methods involving null surfaces
should be formulated in a way that can accomodate surfaces of arbitrary
causal type.\\
\\
The purpose of this paper is thus to fill this gap in the literature
and provide a formulation of thin shells and junction conditions for
GR through a variational principle valid for a generic hypersurface.
A natural question is then why should one consider generic shells.
One reason is that it is beneficial to provide a unified formalism
capable of encompassing timelike, spacelike and null shells at the
same time, rather than assuming the signature from the beginning.
As the example of the Barrabes-Israel formalism shows (remarked in
Footnote \ref{fn:The-work-by}), such unified formulations tend to
include the case of non-pure hypersurfaces as well. Moreover, as argued
before, even if one is interested in null shells exclusively, the
convenient setting for a variational treatment of null hypersurfaces
is the one which is applicable to generic hypersurfaces equally. For
another reason, non-pure hypersurfaces themselves can appear in physically
interesting situations. Some examples may be found in  \cite{Sen}.
To give one explicitly, the stationary limit surface of a Kerr black
hole is timelike almost everywhere but null at a set of measure zero.
If one wishes to obtain matching conditions for spacetime regions
separated by such hypersurfaces, one must incorporate signature-changing
hypersurfaces. For an application of matching non-pure hypersurfaces, we refer to the works \cite{MSV,MSV2,MSV3} by Mars, Senovilla and Vera on signature change on brane worlds.\\
\\
The primary motivation for the development of this work is the formulation of thin shell equations in modified theories of gravitation. Thin shells have already been considered in extended gravitational theories, for example in \cite{Av} thin shells and junction conditions have been examined in Brans-Dicke type scalar-tensor theories via the distributional formalism with the null and non-null cases separately. A variational formalism has also been given but only for the non-null case. In \cite{Dav,Grav}, junction conditions have been formulated in Gauss-Bonnet gravity for applications to Gauss-Bonnet brane worlds via the variational formulation, but once again only for non-null cases. Shells in higher order gravity have also been investigated in \cite{Reina} through the use of distributions. Higher order theories have qualitatively different shell behaviour with so-called double layers - energy-momentum terms proportional to the Dirac delta's derivative - appearing.\\
\\
The most general scalar-tensor theory with second-order field equations is Horndeski's theory (originally published as \cite{Horn}, but the most common form is the equivalent "DGSZ reformulation" \cite{DGSZ}). Thin shell equations in Horndeski's theory have been found by Padilla and Sivanesan \cite{PS} through a variational method valid only for non-null hyper surfaces.\\
\\
In  \cite{RG} we gave a formulation of null shells in a reduced class of Horndeski theories
via the distributional method and the qualitative form of these equations
differed greatly from those obtained by Padilla and Sivanesan. For
a more effective comparision it would have been beneficial to also
follow a variational approach, however no such method was found that
would be valid for generic shells, yet it is a valuable and often-used method for non-null hypersurfaces. It is thus reasonable to first
examine how the variational formalism works for generic shells in
GR before generalizing to more complicated theories.\\
\\
The main obstacle for such a formalism seems to be the lack of
an appropriate variational counterterm for generic boundaries, as
the Gibbons-Hawking-York term is valid only for timelike and spacelike
surfaces. Counterterms valid for null boundaries have been considered by Parattu \etal \cite{Par_null} and extended to piecewise smooth boundaries involving corner terms by Lehner \etal \cite{Leh}. This formalism can be used when the boundary has separate timelike, null and spacelike pieces but does not allow for a unified treatment or for cases when the boundary has null points that do not form an entire segment (for example the null point is isolated or the null points form a line, etc.). An alternative formulation in terms of tetrads have also been given by Jubb \etal \cite{Jubb}, which nonetheless shares the features of the formulation by Lehner \etal in that it is necessary to break the boundary into pieces of pure signatures instead of giving a fully unified treatment. \\
\\
However  a unified counterterm has been provided recently also by
Parattu \etal in  \cite{Par}, which is valid for any boundary
hypersurface rigged with a transversal vector field and reduces to
the Gibbons-Hawking-York term in the appropriate limit. Although the formulation has not been extended to corner terms, we are mainly interested in smooth shells (as in the hypersurface corresponding to the shell being smooth) and therefore this limitation of the formalism does not affect our results. We show that
this counterterm properly regularizes the action at the shell and
the equations derived in eg. Barrabes and Israel \cite{BI}, Mars
and Senovilla \cite{MS} and Senovilla \cite{Sen} via the distributional
method arise as the natural boundary conditions along the hypersurface.
To make contact with the alternative distributional regularization
procedure of Hajicek and Kijowski \cite{HK}, it is also shown
that the singular part of the Lagrangian supported on a generic surface
is proportional to the difference of the counterterm of Parattu \etal and thus it leads to the same variational principle we obtain
by adding the counterterms manually. Finally, we also derive the correct
shell equation via a first order equivalent to the Einstein-Hilbert
action where no regularization is necessary. This is actually a special
case of the regularization by counterterms as such first order equivalents
can be seen as the Einstein-Hilbert action augmented by a different
counterterm.

\paragraph*{Outline of the paper:}

In Section \ref{sec:Rigged-hypersurfaces} we provide a short summary
of the rigged hypersurface formalism which will be used throughout
this paper. In Section \ref{sec:Variational-counterterms}, several
known variational counterterms for the Einstein-Hilbert action are
discussed including the one recently proposed by Parattu \etal \cite{Par}
valid for generic hypersurfaces. Some general properties of these
counterterms are investigated. The main part of the paper is Section
\ref{sec:Variational-formalism-of}, where the dynamics of thin shells
are formulated as a variational principle via three separate regularization
schemes. Variational counterterms are employed in Subsection \ref{subsec:Thin-shell-ct},
distributional regularization is considered in Subsection \ref{subsec:Thin-shell-dist}
and the shell equation is also derived from a first order action without
the need for regularization in Subsection \ref{subsec:Thin-shell-first}.
Some of the longer calculations are given in \ref{sec:Decomposition-of-PI}
and \ref{sec:Decomposition-of-M}.

\paragraph*{Notation:}

The spacetime manifold is $D+1$ dimensional and is denoted $M$.
Coordinates on $M$ are $x^{\mu}$ with the greek indices running
$\mu=0,1,...,D$. $\Sigma$ is a hypersurface in $M$, that is a $D$
dimensional submanifold with coordinates $y^{a}$ with latin indices
$a,b,c,...$ taking the values $1,...,D$. Summation convention
on repeated indices is assumed. The metric tensor in $M$ is $g_{\mu\nu}$,
its determinant is $\mathfrak{g}$ and the volume form determined
by it is
\begin{equation}
\mu_{g}=\sqrt{-\mathfrak{g}}dx^{0}\wedge...\wedge dx^{D}.
\end{equation}
Inner products with respect to the spacetime metric are denoted with
dots in indexless notation, eg. $X\cdot Y=X^{\mu}Y^{\nu}g_{\mu\nu}$.
All manifolds are assumed orientable and oriented.

\section{Rigged hypersurfaces\label{sec:Rigged-hypersurfaces}}

In this section we review the formalism of rigged hypersurfaces, establishing
the notation to be used in the rest of the paper. We refer to the
exposition by Mars and Senovilla \cite{MS} as well as the works \cite{Mar,Mar2} for proofs of the statements
made here. We also recover the limiting cases when the hypersurface
is timelike or spacelike and we derive the null limit.

\subsection{Structures induced by the rigging}

We consider a hypersurface $\Sigma$ in the $D+1$ dimensional manifold
$M$. The hypersurface is given locally by the embedding functions
\begin{equation}
x^{\mu}=\Phi^{\mu}\left(y^{1},...,y^{D}\right),
\end{equation}
where the $y^{a}$ are the intrinsic coordinates of $\Sigma$. The
derivatives
\begin{equation}
e_{a}^{\mu}:=\frac{\partial\Phi^{\mu}}{\partial y^{a}}
\end{equation}
are interepreted as the components of the holonomic coordinate frame
of $\Sigma$, or from a more invariant point of view, the components
of the pushforward and pullback operations between $\Sigma$ and $M$.
Without introducing any extra structure, a vector $v^{\mu}$ defined
at a point $p\in\Sigma$ is tangential if it can be written in the
form $v^{\mu}=v^{a}e_{a}^{\mu}$ for some intrinsic hypersurface vector
$v^{a}$. Then $v^{\mu}$ is the pushforward of $v^{a}$. Thus, it
is possible to decide whether a contravariant vector (and thus a general
contravariant tensor in an index-by-index basis) is tangential or
not. A covector $n_{\mu}$ defined at some point $p\in\Sigma$ is
normal if $n_{\mu}e_{a}^{\mu}=0$, that is it annihilates all tangential
vectors. The space of normal covectors at each point is one dimensional.
Thus it is meaningful to decide if a covariant vector is normal or
not. If $\omega_{\mu}$ is a covariant tensor at some $p\in\Sigma$,
its pullback to $\Sigma$ is the hypersurface covector $\omega_{a}=\omega_{\mu}e_{a}^{\mu}$
(this notion is extended to all covariant tensors index-by-index).\\
\\
The induced metric or first fundamental form on $\Sigma$ is the pullback
\begin{equation}
h_{ab}=g_{\mu\nu}e_{a}^{\mu}e_{b}^{\nu}.
\end{equation}
The point $p\in\Sigma$ is a null point of the hypersurface if and
only if $h_{ab}\left(p\right)$ is a singular matrix. Since we allow
for null points and thus non-invertible induced metrics, we do not
raise or lower latin indices.\\
\\
To proceed, we need to introduce a vector field $\ell^{\mu}$ along
$\Sigma$, which is nowhere tangential (nor zero). We call this a choice
of rigging and the pair $\left(\Sigma,\ell\right)$ is a rigged hypersurface.
The set $\left(\ell,e_{1},...,e_{D}\right)$ is then a frame of $M$
along $\Sigma$. The choice of rigging selects a unique normal covector
field $n_{\mu}$ which satisfies
\begin{equation}
n_{\mu}\ell^{\mu}=1.
\end{equation}
Then the set $\left(n,\vartheta^{1},...,\vartheta^{D}\right)$ is
the dual frame of $\left(\ell,e_{1},...,e_{D}\right)$, where the
covector fields (along $\Sigma$) $\vartheta_{\mu}^{a}$ are uniquely
determined by the duality relations
\begin{equation}
\vartheta_{\mu}^{a}\ell^{\mu}=0,\quad\vartheta_{\mu}^{a}e_{b}^{\mu}=\delta_{b}^{a}.
\end{equation}
Using $\vartheta_{\mu}^{a}$, given a hypersurface covector $\omega_{a}$,
we can create a spacetime covector $\omega_{\mu}=\vartheta_{\mu}^{a}\omega_{a}$
which satisfies $\omega_{a}=e_{a}^{\mu}\omega_{\mu}$ and $\omega_{\mu}\ell^{\mu}=0$.
Likewise, we can project a spacetime vector $v^{\mu}$ into $\Sigma$
as $v_{\parallel}^{a}=v^{\mu}\vartheta_{\mu}^{a}$, and also obtain
a direct projection operator $P_{\ \nu}^{\mu}$ by pushing forward
$v_{\parallel}^{a}$, i.e.
\begin{equation}
v_{\parallel}^{\mu}=v_{\parallel}^{a}e_{a}^{\mu}=v^{\nu}\vartheta_{\nu}^{a}e_{a}^{\mu}=P_{\ \nu}^{\mu}v^{\nu},
\end{equation}
with
\begin{equation}
P_{\ \nu}^{\mu}=e_{a}^{\mu}\vartheta_{\nu}^{a}=\delta_{\nu}^{\mu}-\ell^{\mu}n_{\nu}.
\end{equation}
With respect to these bases, the spacetime metric and inverse metric
can be expressed as
\begin{eqnarray}
g_{\mu\nu} & =\ell^{2}n_{\mu}n_{\nu}+\lambda_{a}\left(n_{\mu}\vartheta_{\nu}^{a}+\vartheta_{\mu}^{a}n_{\nu}\right)+h_{ab}\vartheta_{\mu}^{a}\vartheta_{\nu}^{b},\nonumber \\
g^{\mu\nu} & =n^{2}\ell^{\mu}\ell^{\nu}+\nu^{a}\left(\ell^{\mu}e_{a}^{\nu}+e_{a}^{\mu}\ell^{\nu}\right)+h_{\ast}^{ab}e_{a}^{\mu}e_{b}^{\nu},\label{eq:metricex}
\end{eqnarray}
where the elements that appear here are given explicitly as
\begin{eqnarray}
\ell^{2} & =\ell_{\mu}\ell^{\mu}=\ell\cdot\ell,\quad n^{2}=n_{\mu}n^{\mu}=n\cdot n,\nonumber \\
\lambda_{a} & =\ell_{\mu}e_{a}^{\mu}=\ell\cdot e_{a},\quad\nu^{a}=n^{\mu}\vartheta_{\mu}^{a}=n\cdot\vartheta^{a},\nonumber \\
h_{\ast}^{ab} & =g^{\mu\nu}\vartheta_{\mu}^{a}\vartheta_{\nu}^{b}=\vartheta^{a}\cdot\vartheta^{b}.
\end{eqnarray}
In particular, $h_{\ast}^{ab}$ may be seen as a pseudo-inverse to
$h_{ab}$.\\
\\
The choice of rigging also gives a volume form
\begin{equation}
\mu_{\ell,g}=\rho_{\ell,g}dy^{1}\wedge...\wedge dy^{D}
\end{equation}
on $\Sigma$ with
\begin{equation}
\rho_{\ell,g}=\sqrt{-\mathfrak{g}}\ell^{\mu}\pi_{\mu\mu_{1}...\mu_{D}}e_{1}^{\mu_{1}}...e_{D}^{\mu_{D}},
\end{equation}
where $\pi_{\mu\mu_{1}...\mu_{D}}$ is the $D+1$ dimensional Levi-Civita
symbol. This particular volume element is such that if $\Omega\subseteq M$
is a compact $D+1$ dimensional domain of integration, whose boundary
$\partial\Omega$ is rigged with an outward pointing transversal $\ell^{\mu}$,
Gauss' theorem takes the form
\begin{equation}
\int_{\Omega}\nabla_{\mu}X^{\mu}\,\mu_{g}=\oint_{\partial\Omega}n_{\mu}X^{\mu}\,\mu_{\ell,g},
\end{equation}
where $n_{\mu}$ is the normal adapted to the rigging, i.e. $n_{\mu}\ell^{\mu}=1$. Further properties of the volume form may be found in \cite{MS,Mar2}.\\
\\
Extrinsic curvature-type quantities may be obtained by differentiating
the frame vectors in the tangential directions as
\begin{eqnarray}
\chi_{ab} & =-n_{\nu}e_{a}^{\mu}\nabla_{\mu}e_{b}^{\nu}=e_{a}^{\mu}e_{b}^{\nu}\nabla_{\mu}n_{\nu},\label{eq:chi}\\
\varphi_{a} & =n_{\nu}e_{a}^{\mu}\nabla_{\mu}\ell^{\nu}=-e_{a}^{\mu}\ell^{\nu}\nabla_{\mu}n_{\nu},\\
\psi_{a}^{b} & =e_{a}^{\mu}\vartheta_{\nu}^{b}\nabla_{\mu}\ell^{\nu}=-e_{a}^{\mu}\ell^{\nu}\nabla_{\mu}\vartheta_{\nu}^{b}.
\end{eqnarray}
These are all hypersurface tensors and $\chi_{ab}$ is symmetric.
For thin shells and junction conditions it is also useful to define
\begin{equation}
H_{ab}=e_{a}^{\mu}e_{b}^{\nu}\nabla_{\mu}\ell_{\nu}=\psi_{a}^{c}h_{bc}+\varphi_{a}\lambda_{b},\label{eq:H}
\end{equation}
which is non-symmetric in general and is not independent of the triplet
$\left(\chi,\psi,\varphi\right)$. However, it will turn out that
this quantity is what most naturally generalizes the extrinsic curvature
to thin shells and will play an important role. For $H_{ab}$ we make
an exception to our convention not to raise latin indices and define
\begin{eqnarray}
H_{\ b}^{a} & =h_{\ast}^{ac}H_{cb},\quad H_{a}^{\ b}=h_{\ast}^{bc}H_{ac},\quad H^{ab}=h_{\ast}^{ac}h_{\ast}^{bd}H_{cd},\nonumber \\
H & =H_{a}^{a}=H_{ab}h_{\ast}^{ab}.
\end{eqnarray}
A connection-type quantity $\gamma_{\ ab}^{c}$ is also given on $\Sigma$
by
\begin{equation}
\gamma_{\ ab}^{c}=\vartheta_{\nu}^{c}e_{a}^{\mu}\nabla_{\mu}e_{b}^{\nu}=-e_{a}^{\mu}e_{b}^{\nu}\nabla_{\mu}\vartheta_{\nu}^{c}.
\end{equation}
This connection is torsionless but is not metric compatible in general.

\subsection{Transformations between riggings\label{subsec:Transformation-of-the}}

The choice of rigging $\ell^{\mu}$ along a hypersurface $\Sigma\subseteq M$
is not unique and it may be subjected to two kinds of transformations.
The first is a tangential shift
\begin{equation}
\ell^{\prime\mu}=\ell^{\mu}+T^{a}e_{a}^{\mu},
\end{equation}
where $T^{a}$ is an arbitrary tangent vector field to $\Sigma$,
and the second kind is a rescaling
\begin{equation}
\bar{\ell}^{\mu}=\alpha\ell^{\mu},
\end{equation}
where $\alpha$ is a function along $\Sigma$ that is nowhere vanishing. These transformations form a group parametrized by $D+1$ functions whose structure has been analyzed in \cite{Mar2}.
The quantities associated with rigged hypersurfaces transform under
the shift as \cite{MS,Mar}
\begin{eqnarray}
\ell^{\prime\mu} & =\ell^{\mu}+T^{a}e_{a}^{\mu},\nonumber \\
\vartheta_{\mu}^{\prime a} & =\vartheta_{\mu}^{a}-T^{a}n_{\mu},\nonumber \\
\ell^{\prime2} & =\ell^{2}+2T^{a}\lambda_{a}+h_{ab}T^{a}T^{b},\nonumber \\
\lambda_{a}^{\prime} & =\lambda_{a}+h_{ab}T^{b},\nonumber \\
\nu^{\prime a} & =\nu^{a}-n^{2}T^{a},\nonumber \\
h_{\ast}^{\prime ab} & =h^{ab}-\nu^{a}T^{b}-T^{a}\nu^{b}+n^{2}T^{a}T^{b},\nonumber \\
\varphi_{a}^{\prime} & =\varphi_{a}-\chi_{ab}T^{b},\nonumber \\
\psi_{a}^{\prime b} & =\psi_{a}^{b}+\varphi_{a}T^{b}+D_{a}T^{b}+\chi_{ac}T^{b}T^{c},\nonumber \\
\gamma_{ab}^{\prime c} & =\gamma_{ab}^{c}+\chi_{ab}T^{c},\nonumber \\
H_{ab}^{\prime} & =H_{ab}+D_{a}T^{c}h_{bc}-\chi_{ac}\lambda_{b}T^{c},
\end{eqnarray}
while $e_{a}^{\mu}$, $n_{\mu}$, $h_{ab}$, $\chi_{ab}$ and $\mu_{\ell,g}$
are invariant. Under a rescaling, the transformations are
\begin{eqnarray}
\bar{\ell}^{\mu} & =\alpha\ell^{\mu},\nonumber \\
\bar{n}_{\mu} & =\alpha^{-1}n_{\mu},\nonumber \\
\bar{\ell}^{2} & =\alpha^{2}\ell^{2},\nonumber \\
\bar{n}^{2} & =\alpha^{-1}n^{2},\nonumber \\
\bar{\lambda}_{a} & =\alpha\lambda_{a},\nonumber \\
\bar{\nu}^{a} & =\alpha^{-1}\nu^{a}\nonumber \\
\bar{\chi}_{ab} & =\alpha^{-1}\chi_{ab},\nonumber \\
\bar{\varphi}_{a} & =\varphi_{a}+\partial_{a}\ln\alpha,\nonumber \\
\bar{\psi}_{a}^{b} & =\alpha\psi_{a}^{b},\nonumber \\
\bar{H}_{ab} & =\alpha H_{ab}+\partial_{a}\alpha\lambda_{b},\nonumber \\
\bar{\mu}_{\ell,g} & =\alpha\mu_{\ell,g},
\end{eqnarray}
while $e_{a}^{\mu}$, $\vartheta_{\mu}^{a}$, $h_{ab}$, $h_{\ast}^{ab}$
and $\gamma_{ab}^{c}$ are invariant. Note that since the volume element
$\mu_{\ell,g}$ is invariant under shifts, the definition of $\mu_{\ell,g}$
essentially depends on that of the normal $n_{\mu}$ only. Thus if
one has a preferred normal along a hypersurface, the scaling of the
normal already fixes the volume element without the need to choose
a rigging explicitly.

\subsection{Pseudo-Riemannian limit of rigged hypersurfaces\label{subsec:Pseudo-Riemannian-limit-of}}

The usual formalism of timelike and spacelike (collectively, pseudo-Riemannian)
hypersurfaces may be obtained from the rigged formalism by making
a particular choice of rigging $\ell^{\mu}$. We assume that $\Sigma$
is timelike or spacelike and set
%\epsilon=\pm 1=\cases{+1 &for \Sigma\text{ is timelike} \\ -1 &for \Sigma\text{ is spacelike}}
\begin{equation}
\epsilon=\pm 1=\cases{+1& $\Sigma$ is timelike\\
-1& $\Sigma$ is spacelike\\}
\end{equation}
to allow for both cases to be considered simultaneously. The the induced
metric $h_{ab}$ is nondegenerate throughout $\Sigma$, its inverse
$h^{ab}$ exists and we raise and lower latin indices with $h^{ab}$
and $h_{ab}$ respectively. Normal vectors are everywhere transversal,
therefore we take as the rigging
\begin{equation}
\ell^{\mu}=\hat{n}^{\mu}
\end{equation}
the unit normal (i.e. $\hat{n}\cdot\hat{n}=\epsilon$) to $\Sigma$,
which is unique up to sign. With this particular choice of the rigging,
the normal associated to the rigging is
\begin{equation}
n_{\mu}=\epsilon\hat{n}_{\mu}.
\end{equation}
We will only use $\hat{n}$ and keep track of the $\epsilon$s that
appear. The rest of the quantities become
\begin{eqnarray}
\vartheta_{\mu}^{a} & =e_{\mu}^{a}=h^{ab}g_{\mu\nu}e_{b}^{\nu},\nonumber \\
\ell^{2} & =n^{2}=\epsilon,\nonumber \\
\lambda_{a} & =\nu^{a}=0,\nonumber \\
h_{\ast}^{ab} & =h^{ab},\nonumber \\
\chi_{ab} & =\epsilon K_{ab},\nonumber \\
\varphi_{a} & =0,\nonumber \\
\psi_{a}^{b} & =K_{a}^{b},\nonumber \\
H_{ab} & =K_{ab},\nonumber \\
\mu_{\ell,g} & =\mu_{h}=\sqrt{-\epsilon\mathfrak{h}}dy^{1}\wedge...\wedge dy^{D},
\end{eqnarray}
where
\begin{equation}
K_{ab}=e_{a}^{\mu}e_{b}^{\nu}\nabla_{\mu}\hat{n}_{\nu}=\frac{1}{2}e_{a}^{\mu}e_{b}^{\nu}\mathcal{L}_{\hat{n}}g_{\mu\nu}
\end{equation}
is the usual extrinsic curvature and $\mathfrak{h}=\det\left(h_{ab}\right)$
is the determinant of the induced metric. The connection $\gamma_{\ ab}^{c}$
becomes the Levi-Civita connection of the induced metric $h_{ab}$
with
\begin{equation}
\gamma_{\ ab}^{c}=\frac{1}{2}h^{cd}\left(\partial_{a}h_{bd}+\partial_{b}h_{ad}-\partial_{d}h_{ab}\right)
\end{equation}
and Gauss' theorem takes the form
\begin{equation}
\int_{\Omega}\nabla_{\mu}X^{\mu}\,\mu_{g}=\oint_{\partial\Omega}\epsilon\hat{n}_{\mu}X^{\mu}\,\mu_{h},
\end{equation}
where $\hat{n}^{\mu}$ is the outward pointing unit normal to $\partial\Omega$.

\subsection{Null limit of rigged hypersurfaces\label{subsec:Null-limit-of}}

Suppose now that $\Sigma$ is null. There is no universally preferred
convention here for the rigging, however the null rigging used by
eg. Poisson \cite{Poi} is a useful choice and we present it here.
If $\Sigma$ is null then any normal field $n_{\mu}$ is also null
and is tangential to $\Sigma$. Moreover it satisfies the geodesic
equation
\begin{equation}
\left(\nabla_{n}n\right)^{\mu}=\kappa n^{\mu}
\end{equation}
for some \emph{non-affinity function} $\kappa$. We may then set up
coordinates $\left(y^{a}\right)=\left(r,\theta^{A}\right)$ on $\Sigma$
($A,B,...=2,...,D$) such that
\begin{equation}
n^{\mu}=\left(\frac{\partial}{\partial r}\right)^{\mu},
\end{equation}
and choose a null $\ell^{\mu}$ rigging which satisfies
\begin{equation}
\ell^{\mu}\ell_{\mu}=0,\quad\ell^{\mu}n_{\mu}=1,\quad\ell_{\mu}e_{A}^{\mu}=0,
\end{equation}
where
\begin{equation}
e_{A}^{\mu}=\left(\frac{\partial}{\partial\theta^{A}}\right)^{\mu}
\end{equation}
are the rest of the basis fields, necessarily spacelike. The functions
\begin{equation}
q_{AB}=e_{A}^{\mu}e_{B}^{\nu}g_{\mu\nu}
\end{equation}
are then the components of the spacelike induced metric on the slices
$r=\mathrm{const}$. They are also the only nonvanishing components
of the induced metric on the entire surface, i.e.
\begin{equation}
\left(h_{ab}\right)=\left(\begin{array}{*{20}{c}}0&0\\
0&\left(q_{AB}\right)
\end{array}\right).
\end{equation}
The $D-1$-metric $q_{AB}$ does possess an inverse, denoted $q^{AB}$
and the capital latin indices are raised and lowered via $q^{AB}$
and $q_{AB}$ respectively. The most important extrinsic curvature
quantity in this case is $H_{ab}$, which is now symmetric and we
split it as $H_{11}$, $H_{1A}$ and $H_{AB}$. We have
\begin{eqnarray}
H_{11} & =n^{\mu}n^{\nu}\nabla_{\mu}\ell_{\nu}=-n^{\mu}\ell_{\nu}\nabla_{\mu}n^{\nu}=-\kappa,\nonumber \\
H_{1A} & =H_{A1}=e_{A}^{\mu}n^{\nu}\nabla_{\mu}\ell_{\nu},\nonumber \\
H_{AB} & =e_{A}^{\mu}e_{B}^{\nu}\nabla_{\mu}\ell_{\nu}.
\end{eqnarray}
We may express most of the quantities with $H_{ab}$ as
\begin{eqnarray}
\varphi_{1} & =n^{\mu}n_{\nu}\nabla_{\mu}\ell^{\nu}=-\kappa=H_{11},\nonumber \\
\varphi_{A} & =e_{A}^{\mu}n_{\nu}\nabla_{\mu}\ell^{\nu}=H_{A1},\nonumber \\
\psi_{1}^{1} & =n^{\mu}\ell_{\nu}\nabla_{\mu}\ell^{\nu}=\frac{1}{2}n^{\mu}\nabla_{\mu}\left(\ell_{\nu}\ell^{\nu}\right)=0,\nonumber \\
\psi_{A}^{1} & =e_{A}^{\mu}\ell_{\nu}\nabla_{\mu}\ell^{\nu}=0,\nonumber \\
\psi_{1}^{A} & =n^{\mu}e_{\nu}^{A}\nabla_{\mu}\ell^{\nu}=H_{1}^{A},\nonumber \\
\psi_{A}^{B} & =H_{A}^{B}.
\end{eqnarray}
The primary exception is $\chi_{ab}$, which is
\begin{eqnarray}
\chi_{11} & =n^{\mu}n^{\nu}\nabla_{\mu}n_{\nu}=\kappa n^{\nu}n_{\nu}=0,\nonumber \\
\chi_{1A} & =n^{\mu}e_{A}^{\nu}\nabla_{\mu}n_{\nu}=\kappa e_{A}^{\nu}n_{\nu}=0,\nonumber \\
\chi_{AB} & =e_{A}^{\mu}e_{B}^{\nu}\nabla_{\mu}n_{\nu},
\end{eqnarray}
and is thus not expressible with $H_{ab}$. As only tangential derivatives
of tangential vectors are taken, when thin shells are involved, the
jump of this quantity always vanishes.\\
\\
Finally, with respect to the frame $\left(\ell,n,e_{A}\right)$ the
full metric tensor has components
\begin{equation}
\left(g_{\mu\nu}\right)=\left(\begin{array}{*{20}{c}}0 & 1 & 0\\
1 & 0 & 0\\
0 & 0 & \left(q_{AB}\right)
\end{array}\right),
\end{equation}
from which it follows that in this frame
\begin{equation}
\mathfrak{g}=-\mathfrak{q},
\end{equation}
where $\mathfrak{q}=\det\left(q_{AB}\right)$. The volume element
can be thus written as
\begin{equation}
\mu_{\ell,g}=\mu_{q}=\sqrt{\mathfrak{q}}dr\wedge d\theta^{2}...\wedge d\theta^{D}.
\end{equation}
Note that while it appears that the volume element $\mu_{q}$ is canonically
given, it does depend on the way the manifold $\Sigma$ has been sliced
into spacelike $D-1$-surfaces.

\section{Variational counterterms\label{sec:Variational-counterterms}}

The Einstein-Hilbert action over $M$ is
\begin{equation}
S_{\mathrm{EH}}=\frac{1}{2\varkappa}\int_{M}R\,\mu_{g},
\end{equation}
where $\varkappa=8\pi G$. The integrand is second order in the metric
while its Euler-Lagrange equations are also second order. If we assume
the boundary $\partial M$ has been rigged by an outward pointing
vector $\ell^{\mu}$, we may write its variation in generic form as
\begin{equation}
\fl\delta S_{\mathrm{EH}}=-\int_{M}\frac{1}{2\varkappa}G^{\mu\nu}\delta g_{\mu\nu}\,\mu_{g}+\int_{\partial M}\left(Y^{\mu\nu}\delta g_{\mu\nu}+Y^{\mu\nu,a}\delta g_{\mu\nu,a}+Y_{\ell}^{\mu\nu}\delta g_{\mu\nu,\ell}\right)\mu_{g,\ell},
\end{equation}
where $\delta g_{\mu\nu,a}=e_{a}^{\kappa}\partial_{\kappa}\delta g_{\mu\nu}$
are the tangential derivatives of the metric variation, $\delta g_{\mu\nu,\ell}=\ell^{\kappa}\partial_{\kappa}\delta g_{\mu\nu}$
is the transversal derivative and $Y^{\mu\nu}$, $Y^{\mu\nu,a}$ and
$Y_{\ell}^{\mu\nu}$ are the appropriate coefficients that appear
on the boundary. Imposing Dirichlet boundary conditions $\left.\delta g_{\mu\nu}\right|_{\partial M}=0$
gets rid of the first two terms on the boundary, but not the third.
On the other hand demanding the transversal derivatives to also vanish
would overdetermine the field equations. In order to make the variational
problem well-defined a variational counterterm
\begin{equation}
B=\int_{\partial M}\mathcal{B}\left(g,\left(\partial g\right)_{\parallel},\left(\partial g\right)_{\ell}\right)d^{D}y
\end{equation}
is added to the action, where $\left(\partial g\right)_{\parallel}$
and $\left(\partial g\right)_{\ell}$ are schematic notations for
the tangential and transversal derivatives respectively. If the integrand
$\mathcal{B}$ satisfies
\begin{equation}
\frac{\partial\mathcal{B}}{\partial g_{\mu\nu,\ell}}=-Y_{\ell}^{\mu\nu}\rho_{g,\ell},
\end{equation}
then it follows that imposing the usual Dirichlet condition $\left.\delta g_{\mu\nu}\right|_{\partial M}=0$
on the combined action $S_{\mathrm{EH}}+B$ will get rid of all boundary
terms. The variational counterterm is not unique, however if $\mathcal{B}$
and $\mathcal{B}^{\prime}$ are both integrands of variational counterterms,
their derivatives with respect to $g_{\mu\nu,\ell}$ must be the same
function $-Y_{\ell}^{\mu\nu}\rho_{g,\ell}$ and thus the difference
$\mathcal{B}-\mathcal{B}^{\prime}$ is a function of $g$ and $\left(\partial g\right)_{\parallel}$
only. This result will be of significance for thin shells.\\
\\
There are several counterterms known for the Einstein-Hilbert action:

\paragraph*{The Gibbons-Hawking-York counterterm:}

When the boundary $\partial M$ consists of pseudo-Riemannian pieces
only, the appropriate rigging (see Section \ref{subsec:Pseudo-Riemannian-limit-of})
can be chosen. The counterterm is
\begin{equation}
B_{\mathrm{GHY}}=\frac{\epsilon}{\varkappa}\int_{\partial M}K\,\mu_{h}.
\end{equation}
Its validity follows from the variational formula \cite{Pad}
\begin{equation}
\fl\delta S_{\mathrm{EH}}=-\frac{1}{2\varkappa}\int_{M}G^{\mu\nu}\delta g_{\mu\nu}\,\mu_{g}+\frac{\epsilon}{2\varkappa}\int_{\partial M}\left(Kh^{ab}-K^{ab}\right)\delta h_{ab}\,\mu_{h}-\frac{\epsilon}{\varkappa}\delta\left(\int_{\partial M}K\,\mu_{h}\right).\label{eq:varform}
\end{equation}
The first boundary term involves only tangential derivatives of the
metric and the second term - which contains normal derivatives - is
an exact variation. Adding this term to the action with an opposite
sign will ensure that fixing the metric without fixing its derivatives
on the boundary makes all remaining boundary terms vanish.

\paragraph*{The Einstein counterterm:}

We assume that the manifold $M$ is covered by the domain of a chosen
(and fixed) coordinate chart $x^{\mu}$. Let us also take an outward
pointing normal $n_{\mu}$ along $\partial M$, and let $\mu_{g,n}$
denote the corresponding volume element obtained via \emph{any} rigging
$\ell^{\mu}$ which satisfies $\ell^{\mu}n_{\mu}=1$. The Einstein
counterterm is then defined as
\begin{equation}
B_{\mathrm{E}}=-\frac{1}{2\varkappa}\int_{\partial M}n_{\kappa}w^{\kappa}\,\mu_{g,n},
\end{equation}
where
\begin{equation}
w^{\kappa}=\Gamma_{\ \mu\nu}^{\kappa}g^{\mu\nu}+\Gamma_{\ \nu\mu}^{\nu}g^{\kappa\mu}.\label{eq:50}
\end{equation}
This expression is naturally defined in the interior of $M$ as well,
and by Gauss' theorem
\begin{equation}
B_{E}=-\frac{1}{2\varkappa}\int_{M}\nabla_{\kappa}w^{\kappa}\,\mu_{g},
\end{equation}
where the covariant derivative treats $w^{\kappa}$ as if it was a
vector field (the rationale behind this is that we may consider $\partial_{\mu}$
to be a locally defined connection associated to the chart $x^{\mu}$,
and from this point of view the connection coefficients $\Gamma_{\ \mu\nu}^{\kappa}$
are tensor components - the components of the difference tensor between
$\nabla$ and $\partial$). Decomposing the scalar curvature as
\begin{equation}
R=\nabla_{\kappa}\left(\Gamma_{\ \mu\nu}^{\kappa}g^{\mu\nu}+g^{\kappa\mu}\Gamma_{\ \nu\mu}^{\nu}\right)+\left(\Gamma_{\ \mu\rho}^{\kappa}\Gamma_{\ \kappa\nu}^{\rho}-\Gamma_{\ \kappa\rho}^{\kappa}\Gamma_{\ \mu\nu}^{\rho}\right)g^{\mu\nu},
\end{equation}
one obtains
\begin{equation}
S_{\mathrm{EH}}+B_{\mathrm{E}}=\frac{1}{2\varkappa}\int_{M}g^{\mu\nu}\left(\Gamma_{\ \mu\rho}^{\kappa}\Gamma_{\ \kappa\nu}^{\rho}-\Gamma_{\ \kappa\rho}^{\kappa}\Gamma_{\ \mu\nu}^{\rho}\right)\mu_{g},\label{eq:GG_action}
\end{equation}
which is Einstein's first order, noncovariant $\Gamma\Gamma$-action
\cite{LL,Mo}. Since it is first order, fixing the metric
at the boundary is sufficient to eliminate all boundary terms. The
Einstein counterterm is not unique in the sense that different coordinate
systems will produce different Einstein counterterms, as it is clear
from the lack of covariance of (\ref{eq:50}).

\paragraph*{The background connection counterterm:}

We can also introduce an arbitrary torsionless connection $\bar{\nabla}_{\mu}$.
Quantities calculated from $\bar{\nabla}_{\mu}$ are denoted with
an overbar. Let $n_{\mu}$ be any outward pointing normal to the boundary
$\partial M$ and $\mu_{g,n}$ the associated volume element. The
background connection counterterm is
\begin{equation}
B_{\mathrm{BC}}=-\frac{1}{2\varkappa}\int_{\partial M}\left(n_{\kappa}\Delta_{\ \mu\nu}^{\kappa}g^{\mu\nu}+n^{\mu}\Delta_{\ \nu\mu}^{\nu}\right)\mu_{g,n},
\end{equation}
where $\Delta_{\ \mu\nu}^{\kappa}=\Gamma_{\ \mu\nu}^{\kappa}-\bar{\Gamma}_{\ \mu\nu}^{\kappa}$
is the difference tensor. The Einstein counterterm is reproduced if
$M$ fits into a single coordinate domain and we choose $\bar{\nabla}_{\mu}=\partial_{\mu}$.
The term $\Delta_{\ \mu\nu}^{\kappa}g^{\mu\nu}+\Delta_{\ \nu\mu}^{\nu}g^{\kappa\mu}$
is once again defined on the entire manifold $M$ and after using
Gauss' theorem we get
\begin{equation}
S_{\mathrm{EH}}+B_{\mathrm{BC}}=\frac{1}{2\varkappa}\int_{M}\left[\bar{R}+g^{\mu\nu}\left(\Delta_{\ \mu\rho}^{\kappa}\Delta_{\ \kappa\nu}^{\rho}-\Delta_{\ \kappa\rho}^{\kappa}\Delta_{\ \mu\nu}^{\rho}\right)\mu_{g}\right].\label{eq:bc_action}
\end{equation}
Since $\bar{R}=g^{\mu\nu}\bar{R}_{\mu\nu}$ is the scalar curvature
of the nondynamical background connection $\bar{\nabla}_{\mu}$, this
action is also first order, from which immediately follows that fixing
the metric at the boundary removes all boundary terms. Unlike the
Einstein counterterm, the background connection counterterm is globally
defined and both the counterterm and the resulting first order action
are covariant. However the counterterm and action both contain an
unphysical background field. This counterterm is also non-unique,
as it depends on the connection chosen as the background.

\paragraph*{The rigged counterterm:}

This counterterm has been introduced by Parattu \etal \cite{Par}
as a generalization of the Gibbons-Hawking-York term valid for hypersurfaces
of arbitrary causal type. We fix an outward pointing rigging $\ell^{\mu}$
along the boundary $\partial M$. The rigged counterterm is \cite{Par}
\begin{equation}
B_{\mathrm{R}}=\frac{1}{\varkappa}\int_{\partial M}P_{\ \nu}^{\mu}\nabla_{\mu}n^{\nu}\,\mu_{\ell,g},\label{eq:RCT}
\end{equation}
where $P_{\ \nu}^{\mu}=\delta_{\nu}^{\mu}-\ell^{\mu}n_{\nu}$ is a
tangential projector that removes the $\ell$-directed parts of vectors.
Parattu \etal did not employ the formalism of rigged hypersurfaces,
therefore this counterterm appeared in terms of spacetime, rather
than hypersurface quantities. We rewrite it via $P_{\ \nu}^{\mu}=e_{a}^{\mu}\vartheta_{\nu}^{a}$
and $\vartheta^{\mu a}=\nu^{a}\ell^{\mu}+h_{\ast}^{ab}e_{b}^{\mu}$
as
\begin{eqnarray}
P_{\ \nu}^{\mu}\nabla_{\mu}n^{\nu} & =e_{a}^{\mu}\vartheta_{\nu}^{a}\nabla_{\mu}n^{\nu}=e_{a}^{\mu}\left(h_{\ast}^{ab}e_{b}^{\nu}+\nu^{a}\ell^{\nu}\right)\nabla_{\mu}n_{\nu}\nonumber \\
 & =\chi_{ab}h_{\ast}^{ab}-\varphi_{a}\nu^{a},
\end{eqnarray}
thus the counterterm has the equivalent expression
\begin{equation}
B_{\mathrm{R}}=\frac{1}{\varkappa}\int_{\partial M}\left(\chi_{ab}h_{\ast}^{ab}-\varphi_{a}\nu^{a}\right)\mu_{\ell,g},\label{eq:RCT_proj}
\end{equation}
a form resembling the Gibbons-Hawking-York counterterm with $\chi_{ab}$
and $\varphi_{a}$ playing the role of the extrinsic curvature. From
(\ref{eq:RCT_proj}) it can be seen that when the boundary is
pseudo-Riemannian, choosing $\ell^{\mu}=\hat{n}^{\mu}$ gives (see
Section \ref{subsec:Pseudo-Riemannian-limit-of}) $\varphi_{a}=0$,
$\chi_{ab}=\epsilon K_{ab}$ and $h_{\ast}^{ab}=h^{ab}$, thus the
rigged counterterm reproduces the Gibbons-Hawking-York term in this
limit.\\
\\
In the presence of a boundary $\partial M$ of any causal type, equipped
with a rigging, the variational formula (\ref{eq:varform}) is replaced
by \cite{Par}
\begin{equation}
\delta S_{\mathrm{EH}}=-\frac{1}{2\varkappa}\int_{M}G^{\mu\nu}\delta g_{\mu\nu}\mu_{g}+\frac{1}{2\varkappa}\int_{\partial M}\Pi^{\mu\nu}\delta g_{\mu\nu}\mu_{\ell,g}-\frac{1}{\varkappa}\int_{\partial M}\delta\left(P_{\ \nu}^{\mu}\nabla_{\mu}n^{\nu}\mu_{\ell,g}\right),\label{eq:varformgen}
\end{equation}
where
\begin{equation}
\Pi^{\mu\nu}=g^{\mu\nu}\left(P_{\ \sigma}^{\rho}\nabla_{\rho}n^{\sigma}\right)-\nabla^{(\mu}n^{\nu)}-\nabla_{\rho}\ell^{\rho}n^{\mu}n^{\nu}.\label{eq:pi}
\end{equation}
We remark that this quantity involves the covariant derivative of the normal vector and thus depends on the extension of it to a neighborhood of the boundary. As the calculations in \ref{sec:Decomposition-of-PI} show, the expression is independent of the extension of the normal field. The quantity $\Pi^{\mu\nu}$ will also be decomposed in terms of hypersurface quantities
in Subsection \ref{subsec:Thin-shell-ct}. The sign of the last term
has been corrected as compared to the corresponding result in 
\cite{Par}. The boundary term that results from the variation
of $S_{\mathrm{EH}}+B_{\mathrm{R}}$ is proportional to $\delta g_{\mu\nu}$
and vanishes when the metric is fixed on the boundary. The rigged
counterterm is not unique, different choices of rigging will give
different counterterms.

\section{Variational formalism of thin shells\label{sec:Variational-formalism-of}}

We assume that the hypersurface $\Sigma$ partitions the spacetime
manifold $M$ into two domains $M^{+}$ and $M^{-}$. These domains
are manifolds with boundaries and their interiors are disjoint. For
simplicity we assume that $M$ has no outer boundary, which implies
that $\partial M^{+}=\partial M^{-}=\Sigma$ (this notation currently
ignores orientations). The formalism may be equally well used in the
presence of outer boundaries, but including them would needlessly
complicate the notation and outer boundaries play no role in our formalism
anyway.\\
\\
The regions $M^{+}$ and $M^{-}$ are distinct as manifolds with smooth\footnote{Or is at least $C^{3}$ to ensure both the field equations and the
Bianchi identities exist as regular functions.} metrics $g_{\mu\nu}^{+}$ and $g_{\mu\nu}^{-}$ respectively. Coordinate
systems $x_{+}^{\mu}$ and $x_{-}^{\mu}$ are employed which need
not satisfy any matching conditions at $\Sigma$. As per the analysis
of Clarke and Dray \cite{CD} (also comments made in \cite{MS,Sen}), corrected and extended for the case of hypersurfaces with null points by Mars, Senovilla and Vera \cite{MSV2},
the conditions for the existence of a $C^{1}$ structure on $M$ is
that the induced metrics $h_{ab}^{+}$ and $h_{ab}^{-}$ agree on
$\Sigma$, and in case $\Sigma$ is not timelike or spacelike, there
is a pair of rigging vectors $\ell_{+}^{\mu}$ and $\ell_{-}^{\mu}$
along $\Sigma$ such that $\ell_+$ and $\ell_-$ both point towards $M^+$ (or both towards $M^-$, depending on one's choice) and
\begin{equation}
\lambda_{a}^{+}=\lambda_{a}^{-},\quad\ell_{+}^{2}=\ell_{-}^{2},
\end{equation}
where $\lambda_a^\pm=\ell_\pm\cdot e_a$ are the projections of the transverse vectors $\ell^\mu_\pm$ on the tangent basis of the hypersurface.
This identifies $\ell_{+}$ and $\ell_{-}$ as ``being the same'',
and thus generates a $C^{1}$ differentiable structure at $\Sigma$.
It follows that any coordinate system adapted to the rigging $\ell$
, that is a coordinate system $\left(\sigma,y^{a}\right)$ such that
$\sigma=0$ is the equation for $\Sigma$ and
\begin{equation}
\ell=\frac{\partial}{\partial\sigma},
\end{equation}
is a $C^{1}$ coordinate system. If $\Sigma$ is timelike or spacelike,
then the unit normal $\hat{n}^{\mu}$ always provides a rigging which
satisfies the above conditions, therefore in that case there is no
need to find a pair of matching riggings and it follows that Gaussian
normal coordinates are always $C^{1}$. From this point on we assume
that all spacetime coordinate systems are $C^{1}$ on $\Sigma$ and
$C^{4}$ away from $\Sigma$. Since the final results will be expressed
as hypersurface tensors, this does not reduce the practical applicability
of the formalism. In these coordinate systems, the relation $h_{ab}^{+}=h_{ab}^{-}$
also implies that the spacetime metric $g_{\mu\nu}$ is continuous,
due to expansion (\ref{eq:metricex}), which involves only $\lambda_{a}$,
$\ell^{2}$ and $h_{ab}$, which are then all assumed continuous.\\
\\
We use the notation
\begin{equation}
\left[F\right]=\left.F^{+}\right|_{\Sigma}-\left.F^{-}\right|_{\Sigma}
\end{equation}
for the jump discontinuity of a field $F$ at $\Sigma$ (thus $\left[F\right]$
is a function defined only on $\Sigma$) and
\begin{equation}
\bar{F}=F^{+}\theta+F^{-}\left(1-\theta\right)
\end{equation}
for the ``soldering'' of a field, where
\begin{equation}
\theta(p)=\cases{1& $p\in M^+\setminus\Sigma$\\
0& $p\in M^-\setminus\Sigma$\\
\frac{1}{2}& $p\in\Sigma$}
\label{eq:step}
\end{equation}
is the Heaviside step function associated to $\Sigma$. Any choice of value for $\theta$ at $\Sigma$ ensures that for a continuous
field $F$, $F=\bar{F}$ is a pointwise identity. The choice
$\left.\theta\right|_{\Sigma}=1/2$ is taken for reasons of symmetry. We imagine that the hypersurface of discontinuity $\Sigma$ is the limit of a layer of finite thickness, where the field $F$ is continuous albeit rapidly varying. In the limit of infinitesimal thickness a value between $F^+$ and $F^-$ should be picked on $\Sigma$ and taking the arithmetic average (corresponding to $\left.\theta\right|_\Sigma=1/2$) is the most "democratic" choice that gives no preference to the field values on either side of the layer.\\
\\
To conform to the usual conventions, we also assume that the rigging
vector field $\ell$ points from $M^{-}$ to $M^{+}$. The orientation
on $\Sigma$ is induced by the rigging $\ell$. It follows that $\Sigma$
has the boundary orientation inherited as the boundary of $M^{-}$
and the opposite to the boundary orientation inherited from $M^{+}$.

\subsection{Thin shell equation from the action regularized by counterterms\label{subsec:Thin-shell-ct}}

The total action will be taken to consist of the gravitational action
$S_{\mathrm{EH}}$, an unspecified bulk matter action $S_{\mathrm{M}}$
and an unspecified thin shell matter action $S_{\mathrm{TS}}$. Instead
of integrating over $M$ at once, we split the integrals into sums
of integrals over $M^{+}$ and $M^{-}$. Since $\Sigma$ is not a
part of the outer boundary of the manifold, the usual Dirichlet conditions
do not apply to $\Sigma$, the metric is not fixed there. We suppose
the metric is $C^{0}$ across $\Sigma$ and at least $C^{3}$ away
from $\Sigma$. Since we are varying within this differentiability
class, $\delta g_{\mu\nu}$ also inherits these properties. The equations
of motion of the shell arise as the natural boundary conditions on
the shell as the bulk and boundary contributions to the variation
of the action must vanish separately.\\
\\
The shell hypersurface $\Sigma$ is an interior boundary and thus
we add the rigged counterterm (\ref{eq:RCT}) to the action at both
sides of $\Sigma$ to ensure the proper boundary behaviour of the
action. The total action is then
\begin{eqnarray}
S & =\underbrace{\frac{1}{2\varkappa}\int_{M^{+}}R^{+}\mu_{g}}_{S_{\mathrm{EH}}^{+}}+\underbrace{\frac{1}{2\varkappa}\int_{M^{-}}R^{-}\mu_{g}}_{S_{\mathrm{EH}}^{-}}+\underbrace{\int_{M^{+}}\mathcal{L}_{\mathrm{M}}^{+}d^{D+1}x}_{S_{\mathrm{M}}^{+}}+\underbrace{\int_{M^{-}}\mathcal{L}_{\mathrm{M}}^{-}d^{D+1}x}_{S_{\mathrm{M}}^{-}}\nonumber \\
 & \underbrace{-\frac{1}{\varkappa}\int_{\Sigma}\left(P_{\ \nu}^{\mu}\nabla_{\mu}n^{\nu}\right)_{+}\mu_{g,\ell}}_{B_{\mathrm{R}}^{+}}+\underbrace{\frac{1}{\varkappa}\int_{\Sigma}\left(P_{\ \nu}^{\mu}\nabla_{\mu}n^{\nu}\right)_{-}\mu_{g,\ell}}_{B_{\mathrm{R}}^{-}}+\underbrace{\int_{\Sigma}\mathcal{L}_{\mathrm{TS}}d^{D}y}_{S_{\mathrm{TS}}}.\label{eq:total_action}
\end{eqnarray}
The relative sign difference between $B_{\mathrm{R}}^{+}$ and $B_{\mathrm{R}}^{-}$
is caused by the orientation of $\Sigma$ being opposite to the boundary
orientation inherited from the domain $M^{+}$. Variation of this
integral with respect to the metric is carried out by applying the
variation formula (\ref{eq:varformgen}) to both the $+$ and $-$
integrals, giving
\begin{eqnarray}
\delta S & =\int_{M^{+}}\frac{1}{2}\left(T_{+}^{\mu\nu}-\frac{1}{\varkappa}G_{+}^{\mu\nu}\right)\delta g_{\mu\nu}\,\mu_{g}+\int_{M^{-}}\frac{1}{2}\left(T_{-}^{\mu\nu}-\frac{1}{\varkappa}G_{-}^{\mu\nu}\right)\delta g_{\mu\nu}\,\mu_{g}\nonumber \\
 & +\int_{\Sigma}\left(\frac{1}{\rho_{\ell,g}}\frac{\delta S_{\mathrm{TS}}}{\delta g_{\mu\nu}}-n_{\kappa}\left[\frac{\partial L_{M}}{\partial g_{\mu\nu,\kappa}}\right]-\frac{1}{2\varkappa}\left[\Pi^{\mu\nu}\right]\right)\delta g_{\mu\nu}\,\mu_{\ell,g},
\end{eqnarray}
where $L_{M}=\mathcal{L}_{\mathrm{M}}/\sqrt{-\mathfrak{g}}$ is the
scalarized matter Lagrangian. The variation of the integral should
vanish for all variations $\delta g_{\mu\nu}$ that are $C^{0}$ across
$\Sigma$ and $C^{3}$ away from $\Sigma$. In particular, we can
choose an arbitrary $\delta g_{\mu\nu}$ which satisfies $\left.\delta g_{\mu\nu}\right|_{\Sigma}=0$,
which implies that the coefficients of the $\delta g_{\mu\nu}$ in
the bulk integrals should vanish, giving the Einstein field equations
in the bulk. It then follows that the surface term $\int_{\Sigma}\left(\cdots\right)\delta g_{\mu\nu}\,\mu_{\ell,g}$
should vanish separately even for arbitrary $\delta g_{\mu\nu}$,
which results in the equation
\begin{equation}
S^{\mu\nu}=\frac{1}{\varkappa}\left[\Pi^{\mu\nu}\right],\label{eq:shelleq_abstract}
\end{equation}
where
\begin{equation}
S^{\mu\nu}=\frac{2}{\rho_{\ell,g}}\frac{\delta S_{\mathrm{TS}}}{\delta g_{\mu\nu}}-2n_{\kappa}\left[\frac{\partial L_{\mathrm{M}}}{\partial g_{\mu\nu,\kappa}}\right]\label{eq:ssem_def}
\end{equation}
is the surface energy-momentum tensor. The second term is a contribution
coming from the bulk matter Lagrangian if it also depends on the derivatives
of the metric tensor, usually via the connection. It arises precisely as follows. If $S_{\mathrm M}$ is the matter action with $S_\mathrm M=\int_M\mathcal L_{\mathrm M}\left(g,\partial g,\psi,\partial\psi\right)d^{D+1}x$, and scalar Lagrangian function $L_\mathrm M=\mathcal L_{\mathrm M}/\sqrt{-\mathfrak g}$, the variation of the matter action with respect to the metric is
\begin{equation}
\delta S_{\mathrm M}=\int_M\left(\left(\frac{\partial\mathcal L_{\mathrm M}}{\partial g_{\mu\nu}}-\partial_\kappa\frac{\partial\mathcal L_{\mathrm M}}{\partial g_{\mu\nu,\kappa}}\right)\delta g_{\mu\nu}+\partial_\kappa\left(\frac{\partial\mathcal L_{\mathrm M}}{\partial g_{\mu\nu,\kappa}}\delta g_{\mu\nu}\right)\right)d^{D+1}x.
\end{equation}The total divergence term here can be expressed in terms of the scalar Lagrangian and the covariant divergence. Specifically, since the metric determinant is independent of the metric's first derivative, we get
\begin{equation}
\frac{\partial\mathcal L_{\mathrm M}}{\partial g_{\mu\nu,\kappa}}=\frac{\partial L_{\mathrm M}}{\partial g_{\mu\nu,\kappa}}\sqrt{-\mathfrak g},
\end{equation}
and even though $g_{\mu\nu,\kappa}$ is not a tensor, $\partial L_{\mathrm M}/\partial g_{\mu\nu,\kappa}$ is (see \cite{Schouten}, Chapter II, §11). We can therefore write
\begin{eqnarray}
\delta S_{\mathrm M}&=\mathrm{Bulk\ terms}+\int_M\nabla_\kappa\left(\frac{\partial L_{\mathrm M}}{\partial{g_{\mu\nu,\kappa}}}\delta g_{\mu\nu}\right)\mu_g\nonumber \\ &=\mathrm{Bulk\ terms}+\oint_{\partial M}n_\kappa\frac{\partial L_{\mathrm M}}{\partial{g_{\mu\nu,\kappa}}}\delta g_{\mu\nu}\mu_{\ell,g}.
\end{eqnarray}
If this integral is performed over a spacetime with a shell $\Sigma$ we thus obtain the difference term $-n_\kappa\left[\frac{\partial L_{\mathrm M}}{\partial{g_{\mu\nu,\kappa}}}\right]$ on the shell which contributes to the energy-momentum tensor.
Out of the standard
model fields, only the Lagrangian of the Dirac field depends on the
connection, however the Dirac field being spinorial, an alternative
formulation based on orthonormal tetrads would be necessary to incorporate
them into the formalism. For some exotic matter fields (for example
the scalar sector of Horndeski's theory \cite{Horn}) this term
may be nonvanishing. As far as we are aware, such possible contributions
to the thin shell energy-momentum tensor have not been explored so
far in the literature.\\
\\
\Eref{eq:shelleq_abstract} is the equation of motion for the
thin shell in unprojected form. To proceed, we decompose the tensor
$\Pi^{\mu\nu}$ in the frame $\left(\ell,e_{a}\right)$. This is best
accomplished by transitioning to a coordinate system $\left(\sigma,y^{a}\right)$
adapted to the rigging $\ell$ (i.e. $\ell^{\mu}=\left(\partial/\partial\sigma\right)^{\mu}$and
the $y^{a}$ are the hypersurface coordinates), giving
\begin{eqnarray}
\Pi^{00} & =\chi_{ab}\left(n^{2}h_{\ast}^{ab}-\nu^{a}\nu^{b}\right)+n^{2}\varphi_{a}\nu^{a}-\left(n^{2}\right)^{2}\psi,\nonumber \\
\Pi^{0a} & =\chi_{cd}\left(\nu^{a}h_{\ast}^{cd}-\nu^{c}h_{\ast}^{ad}\right)+n^{2}\varphi_{c}h_{\ast}^{ac}-n^{2}\psi\nu^{a},\nonumber \\
\Pi^{ab} & =\chi_{cd}\left(h_{\ast}^{ab}h_{\ast}^{cd}-h_{\ast}^{ac}h_{\ast}^{bd}\right)+\varphi_{c}\left(\nu^{a}h_{\ast}^{bc}-h_{\ast}^{ab}\nu^{c}+h_{\ast}^{ac}\nu^{b}\right)-\psi\nu^{a}\nu^{b},\label{eq:pi_projected}
\end{eqnarray}
where $\psi=\psi_{a}^{a}$ is the trace. The details of this derivation
may be found in \ref{sec:Decomposition-of-PI}.\\
\\
Since the metric is continuous, only the extrinsic curvature-type
quantities $\chi_{ab}$, $\varphi_{a}$, $\psi_{a}^{b}$ may suffer
jumps, as they in general involve the metric's transversal derivatives.
The reason for the introduction of the tensor $H_{ab}$ in (\ref{eq:H})
has been that as it turns out the jumps of all such quantities may
be related to that of $H_{ab}$. We refer to Mars and Senovilla for
details (equations. (72-76) in  \cite{MS}) and merely list the jump
relations
\begin{eqnarray}
\left[\psi_{a}^{b}\right] & =\left[H_{ac}\right]h_{\ast}^{bc},\nonumber \\
\left[\varphi_{a}\right] & =\left[H_{ab}\right]\nu^{b},\nonumber \\
\left[\chi_{ab}\right] & =n^{2}\left[H_{ab}\right],\nonumber \\
\left[\gamma_{\ ab}^{c}\right] & =-\left[H_{ab}\right]\nu^{c},\label{eq:jump_rel}
\end{eqnarray}
where
\begin{eqnarray}
\left[H_{ab}\right] & =e_{a}^{\mu}e_{b}^{\nu}\left[\nabla_{\mu}\ell_{\nu}\right]=-e_{a}^{\mu}e_{b}^{\nu}\left[\Gamma_{\ \mu\nu}^{\kappa}\right]\ell_{\kappa},
\end{eqnarray}
and is always symmetric. The jump of the metric derivatives can be
written as
\begin{eqnarray}
\left[\partial_{\kappa}g_{\mu\nu}\right] & =\delta_{\kappa}^{\lambda}\left[\partial_{\lambda}g_{\mu\nu}\right]=\left(e_{a}^{\lambda}\vartheta_{\kappa}^{a}+\ell^{\lambda}n_{\kappa}\right)\left[\partial_{\lambda}g_{\mu\nu}\right]\nonumber \\
 & =\left[\partial_{a}g_{\mu\nu}\right]\vartheta_{\kappa}^{a}+\left[g_{\mu\nu,\ell}\right]n_{\kappa}=\left[g_{\mu\nu,\ell}\right]n_{\kappa},\label{eq:dg_jump}
\end{eqnarray}
where the jump of the tangential derivative $\left[\partial_{a}g_{\mu\nu}\right]\vartheta_{\kappa}^{a}$
vanishes because of the continuity of the metric and $g_{\mu\nu,\ell}=\ell^{\kappa}\partial_{\kappa}g_{\mu\nu}$
is the transversal derivative. We then have
\begin{equation}
\left[\Gamma_{\ \mu\nu}^{\kappa}\right]=\frac{1}{2}\left(n_{\mu}\xi_{\nu}^{\kappa}+n_{\nu}\xi_{\mu}^{\kappa}-n^{\kappa}\xi_{\mu\nu}\right),\label{eq:conn_jump}
\end{equation}
where $\xi_{\mu\nu}:=\left[g_{\mu\nu,\ell}\right]$ and it follows
that
\begin{equation}
\left[H_{ab}\right]=\frac{1}{2}e_{a}^{\mu}e_{b}^{\nu}\xi_{\mu\nu}=\frac{1}{2}e_{a}^{\mu}e_{b}^{\nu}\left[g_{\mu\nu,\ell}\right],
\end{equation}
which in adapted coordinates is the jump of the transversal derivative
of the induced metric,
\begin{equation}
\left[H_{ab}\right]=\frac{1}{2}\left[\frac{\partial h_{ab}}{\partial\sigma}\right].
\end{equation}
For this reason it is $\left[H_{ab}\right]$ that carries information
about the discontinuities of the metric's transversal development.\\
\\
Inserting the jump relations (\ref{eq:jump_rel}) into (\ref{eq:pi_projected})
gives
\begin{eqnarray}
\left[\Pi^{00}\right] & =\left[\chi_{ab}\right]\left(n^{2}h_{\ast}^{ab}-\nu^{a}\nu^{b}\right)+n^{2}\left[\varphi_{a}\right]\nu^{a}-\left(n^{2}\right)^{2}\left[\psi\right]\nonumber \\
 & =\left(n^{2}\right)^{2}\left[H\right]-\left(n^{2}\right)^{2}\left[H\right]+n^{2}\left[H_{ab}\right]\nu^{a}\nu^{b}-n^{2}\left[H_{ab}\right]\nu^{a}\nu^{b}\nonumber \\
 & =0,
\end{eqnarray}
\begin{eqnarray}
\left[\Pi^{0a}\right] & =\left[\chi_{cd}\right]\left(\nu^{a}h_{\ast}^{cd}-\nu^{c}h_{\ast}^{ad}\right)+n^{2}\left[\varphi_{c}\right]h_{\ast}^{ac}-n^{2}\left[\psi\right]\nu^{a}\nonumber \\
 & =n^{2}\left[H\right]\nu^{a}-n^{2}\left[H_{cd}\right]\nu^{c}h_{\ast}^{ad}+n^{2}\left[H_{cd}\right]h_{\ast}^{ac}\nu^{d}-n^{2}\left[H\right]\nu^{a}\nonumber \\
 & =0,
\end{eqnarray}
and
\begin{eqnarray}
\left[\Pi^{ab}\right] & =\left[\chi_{cd}\right]\left(h_{\ast}^{ab}h_{\ast}^{cd}-h_{\ast}^{ac}h_{\ast}^{bd}\right)+\left[\varphi_{c}\right]\left(\nu^{a}h_{\ast}^{bc}-h_{\ast}^{ab}\nu^{c}+h_{\ast}^{ac}\nu^{b}\right)-\left[\psi\right]\nu^{a}\nu^{b}\nonumber \\
 & =n^{2}\left(\left[H\right]h_{\ast}^{ab}-\left[H^{ab}\right]\right)+\left[H_{c}^{b}\right]\nu^{a}\nu^{c}+\left[H_{c}^{a}\right]\nu^{b}\nu^{c}\nonumber \\ &-\left[H_{cd}\right]h_{\ast}^{ab}\nu^{c}\nu^{d}-\left[H\right]\nu^{a}\nu^{b}.\label{eq:Pi_jump_proj}
\end{eqnarray}
It follows that the jump $\left[\Pi^{\mu\nu}\right]$ is a tangential
tensor field along $\Sigma$, which we may write as $\left[\Pi^{\mu\nu}\right]=\left[\Pi^{ab}\right]e_{a}^{\mu}e_{b}^{\nu}$.
Following from (\ref{eq:shelleq_abstract}), the surface energy-momentum
tensor must also be tangential with $S^{\mu\nu}=S^{ab}e_{a}^{\mu}e_{b}^{\nu}$
and the shell equation can be considered as the hypersurface tensor
equation
\begin{equation}
\fl\varkappa S^{ab}=n^{2}\left(\left[H\right]h_{\ast}^{ab}-\left[H^{ab}\right]\right)+\left[H_{c}^{b}\right]\nu^{a}\nu^{c}+\left[H_{c}^{a}\right]\nu^{b}\nu^{c}-\left[H_{cd}\right]h_{\ast}^{ab}\nu^{c}\nu^{d}-\left[H\right]\nu^{a}\nu^{b}.\label{eq:shelleq_final}
\end{equation}
Since a contravariant tensor being tangential is an intrinsic notion
independent of any choice of rigging, the components $\left[\Pi^{ab}\right]$
are calculated from $\left[\Pi^{\mu\nu}\right]$ in a way that is
independent of the rigging. Applying the transformation formulae of
Section \ref{subsec:Transformation-of-the} to (\ref{eq:shelleq_final})
shows that $\left[\Pi^{ab}\right]$ is invariant under the shift transformation
$\ell^{\mu}\mapsto\ell^{\mu}+T^{a}e_{a}^{\mu}$ of the rigging and
changes as $\left[\Pi^{ab}\right]\mapsto\alpha^{-1}\left[\Pi^{ab}\right]$
under the rescaling $\ell^{\mu}\mapsto\alpha\ell^{\mu}$. This ambiguity
in the shell equation is related to the fact that for a generic hypersurface
there is no preferred scaling for the normal field $n_{\mu}$. In
the variational principle, both $\left[\Pi^{ab}\right]$ and $S^{ab}$
appear as a factor in the expression
\begin{equation}
\left(S^{ab}-\frac{1}{\varkappa}\left[\Pi^{ab}\right]\right)\delta h_{ab}\,\mu_{\ell,g},
\end{equation}
and the volume element $\mu_{\ell,g}$ depends on the scaling of the
normal. It follows that for the densitized surface tensor $\mathfrak{S}^{ab}=S^{ab}\rho_{\ell,g}$
and densitized $\Pi$-tensor $\mathfrak{P}^{ab}=\Pi^{ab}\rho_{\ell,g}$,
the analogous equation
\begin{equation}
\varkappa\mathfrak{S}^{ab}=\left[\mathfrak{P}^{ab}\right]
\end{equation}
is completely independent of any gauge choices, including the scaling
of the normal. If one wishes to use tensor equations, the scaling
ambiguity in the generic case is unavoidable. For timelike or spacelike
hypersurfaces a canonical choice is given by the unit normal which
fixes the scaling of $\left[\Pi^{ab}\right]$ and $S^{ab}$, while
in the null case Poisson \cite{Poi} gave a physical interpretation
of this ambiguity in terms of observers taking measurements of the
null shell.\\
\\
The tensor $\Pi^{\mu\nu}$ which has been split into the components
$\Pi^{00}$, $\Pi^{0a}$ and $\Pi^{ab}$ may be identified with the
canonical momentum of the gravitational field, up to scaling and densitization
(canonical momenta are usually taken to be tensor densities). Ordinarily,
canonical momenta are constructed by foliating spacetime into a one-parameter
family of spacelike hypersurfaces \cite{Wald}, but one may equally
well consider the analysis of dynamics decomposed with respect to
any foliation of spacetime, including the case when foliate with respect
to the transversal coordinate $\sigma$ adapted to the rigging $\ell^{\mu}$.
In the usual formulation, the canonical momentum is the derivative
of the Lagrangian with respect to ``time'' (which in this case is
$\sigma$) however it is well-known \cite{Nair} that the canonical
momentum may also be identified with the coefficients of the field
variation on the boundary when the Dirichlet conditions are not imposed.
This is the basis for the so-called covariant phase space formalism
\cite{CW,Zuck}. For the Einstein-Hilbert action extended with the
rigged boundary term, by (\ref{eq:varformgen}), the boundary
part is
\begin{equation}
\frac{1}{2\varkappa}\int_{\partial M}\Pi^{\mu\nu}\delta g_{\mu\nu}\mu_{\ell,g}=\frac{1}{2\varkappa}\int_{\partial M}\left(\Pi^{00}\delta\ell^{2}+\Pi^{0a}\delta\lambda_{a}+\Pi^{ab}\delta h_{ab}\right)\mu_{\ell,g},
\end{equation}
which shows that $\Pi^{00}$, $\Pi^{0a}$ and $\Pi^{ab}$ are proportional
to the canonical momenta corresponding to the metric degrees of freedom
$\ell^{2}$, $\lambda_{a}$ and $h_{ab}$. The shell equation then
has the interpretation that the surface energy-momentum tensor is
the jump of the canonical momentum on the hypersuface.\\
\\
If the condition $\left.\delta g_{\mu\nu}\right|_{\partial M}=0$
is not imposed on a boundary (such is the case for thin shells), the
vanishing of the variation of the action forces the coefficients of
the $\delta g_{\mu\nu}$ to vanish on the boundary. Since these coefficients
are identified with the canonical momentum of the field, the canonical
momentum must vanish on the boundary. This is referred to as the natural
boundary condition \cite{GH}, as it arises without having to impose
a boundary condition by hand. We can thus also see that the shell
equation $\frac{1}{2}S^{ab}-\frac{1}{2\varkappa}\left[\Pi^{ab}\right]=0$
is the natural boundary condition for the combined gravitation + bulk
matter + shell matter actions on the hypersurface.\\
\\
Unlike the equations of motions, canonical momenta are not invariant
under equivalence transformations of Lagrangians such as adding total
divergences and - in the case of Einstein-Hilbert type Lagrangians
- they are sensitive to the specific form of the variational counterterm
added to the action. However as discussed in Section \ref{sec:Variational-counterterms},
the difference of two variational counterterms may depend only on
the metric tensor and its tangential derivatives, but never on the
transversal derivative. Only the transversal derivative has nonzero
jump, thus while the expressions $\Pi^{00}$, $\Pi^{0a}$ and $\Pi^{ab}$
depend on the choice of counterterm, their jumps (of which only $\left[\Pi^{ab}\right]$
is nonvanishing) do not. Therefore, the thin shell equation (\ref{eq:shelleq_final})
is actually independent of the choice of counterterm.\\
\\
If $\Sigma$ is timelike or spacelike and we apply the canonical choice
of rigging presented in Section \ref{subsec:Pseudo-Riemannian-limit-of},
we obtain the equation
\begin{equation}
\varkappa S^{ab}=\epsilon\left(\left[K\right]h^{ab}-\left[K^{ab}\right]\right),
\end{equation}
which is the well-known Lanczos equation \cite{Israel}. If instead
we take $\Sigma$ to be null and choose the null rigging adapted to
a spacelike foliation of $\Sigma$ (Section \ref{subsec:Null-limit-of}),
we decompose the equation into components $S^{11}$, $S^{1A}$ and
$S^{AB}$, which are respectively
\begin{eqnarray}
\varkappa S^{11} & =-\left[H_{AB}\right]q^{AB},\nonumber \\
\varkappa S^{1A} & =\left[H_{1B}\right]q^{AB},\nonumber \\
\varkappa S^{AB} & =-\left[H_{11}\right]q^{AB}.
\end{eqnarray}
These relations agree with those of Poisson \cite{Poi}, who interprets
$\mu:=S^{11}$ as the surface energy density, $j^{A}:=S^{1A}$ as
the surface current and - since $S^{AB}$ is diagonal in that it is
proportional to the metric - $p:=-\left[H_{11}\right]$ as the isotropic
surface pressure of the null shell.\\
\\
We conclude this section by comparing the result (\ref{eq:shelleq_final})
to the analogous results in previous works. As mentioned in Footnote
\ref{fn:The-work-by}, Barrabes and Israel \cite{BI} assume $n\cdot n=\mathrm{const}$,
which formally excludes causality-changing hypersurfaces and they
use the normalization $n\cdot\ell=\eta^{-1}$, where $\eta$ is a
nowhere vanishing function along $\Sigma$. One this differing normalization
convention is taken into account, equation (31) in \cite{BI} agrees
with our shell equation (\ref{eq:shelleq_final}). In place of $H_{ab}$,
they employ a different quantity (denoted $\mathcal{K}_{ab}$), the
jump of which however coincides with that of $H_{ab}$ in all cases.\\
\\
In \cite{MS} Mars and Senovilla consider only junction conditions
and analyze the distributional forms of curvature tensors, therefore
the shell equation itself does not appear directly. However since
the energy-momentum tensor is proportional to the Einstein tensor,
the singular part of the Einstein tensor (equation (71) in \cite{MS})
agrees with our $\left[\Pi^{ab}\right]$ up to the appropriate constant
factor and projection. This singular part of the Einstein tensor also
appears in explicitly projected form in equation (23) of \cite{Sen}.

\subsection{Thin shell equation from the action regularized distributionally\label{subsec:Thin-shell-dist}}

Here we explore a different method of regularizing the action integral
at the shell. In the timelike case this method was applied by Hajicek
and Kijowski \cite{HK}. We show that it also works for shells
of any signature. We can write the metric tensor as
\begin{equation}
g_{\mu\nu}=\bar{g}_{\mu\nu}=g_{\mu\nu}^{+}\theta+g_{\mu\nu}^{-}\left(1-\theta\right),
\end{equation}
where $\theta$ is the Heaviside step function defined in (\ref{eq:step}).
This relation is then interpreted distributionally. Reasonably rigorous
treatments of tensor distribution theory on manifolds, can be found
in \cite{dR,Taub,GT,MS,LF}, therefore we only do  here a short review.\\ \\
If $T$ is a type $(k,l)$ tensor field on $M$ we say that a type $(l,k)$ tensor density $\varphi$ of weight $1$ is a \emph{dual density} to $T$, since then the contraction $\langle\varphi,T\rangle=\varphi_{\mu_1...\mu_k}{}^{\nu_1...\nu_l}T^{\mu_1...\mu_k}{}_{\nu_1...\nu_l}$ is a scalar density of weight $1$ that may be integrated over $D+1$ dimensional regions of $M$. Let us define the vector space $D_{k,l}(M)$ to consist of smooth compactly supported tensor densities of type $(l,k)$ (called \emph{test densities}), and the space $D^\ast_{k,l}(M)$ to consist of linear functionals on $D_{k,l}(M)$ that are continuous in the following sense. A linear functional $\chi:D_{k,l}(M)\rightarrow \mathbb R$ is continuous and thus belongs to $D^\ast_{k,l}(M)$ if for each sequence $\varphi_n\in D_{k,l}(M)$ of test densities whose supports are contained in a common compact set $K\subseteq M$ which is itself located in the domain of a coordinate chart, and such that the components $\left(\varphi_n\right)_{\mu_1...\mu_k}{}^{\nu_1...\nu_l}$ and their partial derivatives of all orders tend to $0$ uniformly, we have $\lim_{n\rightarrow \infty}\chi[\varphi_n]=0$. Elements of $D^\ast_{k,l}(M)$ are called \emph{tensor distributions} of type $(k,l)$. A tensor distribution $\chi\in D^\ast_{k,l}(M)$ is \emph{regular} if there exists a (locally integrable but otherwise "rough") tensor field also denoted $\chi$ such that for any test density $\varphi$ we have $\chi[\varphi]=\int_M\langle\chi,\varphi\rangle d^{D+1}x$. This integral converges because $\varphi$ has compact support and since the integrand is a density, no volume form is necessary here. Otherwise the distribution is \emph{singular}. We remark that it is well-defined to take the tensor product of a tensor distribution with a \emph{smooth} tensor field, however products with non-smooth tensor fields only make sense in limited circumstances. \\ \\ de Rham \cite{dR} refers to a distributional $k$-form in the above sense as a \emph{current of degree} $k$ or a $k$-current for short. Since antisymmetric contravariant tensor densities with $(D+1)-k$ indices can be identified canonically with $k$-forms, it follows that the dual densities $\varphi$ of $k$-forms $\omega$ can be canonically identified with $(D+1)-k$-forms under the pairing map $\langle \varphi,\omega\rangle=\omega\wedge\varphi$, thus $k$-currents are continuous linear functionals on $(D+1)-k$-forms. de Rham defines the \emph{boundary} $\partial\omega$ of a $k$-current $\omega$ by $\partial\omega[\varphi]=\omega[d\varphi]$, then the (distributional) exterior derivative by $d\omega=(-1)^{k+1}\partial\omega$. \\ \\ Finally, a few remarks on notation and local representations are in order. As de Rham proves\footnote{de Rham deals only with currents in \cite{dR}, not general tensor distributions. However his arguments are straightforward to generalize to tensor distributions, in fact to distributions modelled on sections of arbitrary vector bundles.} in \cite{dR}, distributions have the sheaf property, i.e. if $D^\ast_{k,l}(U)$ denotes the space of type $(k,l)$ tensor distributions over the open set $U$, and $V\subseteq U$ is an open subset, we have a well-defined restriction map $\mathrm{res}_{V,U}(\chi)\equiv \left.\chi\right|_V$ given by
\begin{equation}
\left.\chi\right|_V[\varphi]:=\chi[\mathrm{ext}_{U,V}(\varphi)],
\end{equation}where $\mathrm{ext}_{U,V}:D_{k,l}(V)\rightarrow D_{k,l}(U)$ extends the tensor density $\varphi\in D_{k,l}(V)$ defined on $V$ with compact support to a tensor density defined on $U$ with compact support by taking $\varphi$ to be zero on $U\setminus V$. This means that the rule $U\mapsto D^\ast_{k,l}(U)$ is a presheaf of real vector spaces, and is in fact a sheaf, i.e. if a tensor distribution vanishes in a neighborhood of each point in $U$, then it vanishes on $U$, and if compatible local distributions are given on an open cover, they glue together to give a well-defined tensor distribution on the covered domain. One may then show that if $U\subseteq M$ is a coordinate domain and $\chi\in D^\ast_{k,l}(U)$ is a tensor distribution of type $(k,l)$, we can write $\chi$ uniquely as
\begin{equation}
\chi=\chi^{\mu_1...\mu_k}{}_{\nu_1...\nu_l}\partial_{\mu_1}\otimes\dots\partial_{\mu_k}\otimes dx^{\nu_1}\otimes\dots\otimes dx^{\nu_l},
\end{equation}where the components are scalar distributions on $U$, and for distributions defined on $M$, the entire distribution may be reconstructed from its sets of components if the manifold is covered by coordinate domains. Moreover, on any test density $\varphi$ we have
\begin{equation}
\chi[\varphi]=\chi^{\mu_1...\mu_k}{}_{\nu_1...\nu_l}\varphi_{\mu_1...\mu_k}{}^{\nu_1...\nu_l}[1],
\end{equation}where the contraction is a distributional scalar density (i.e. $D+1$-form) interpreted as a $D+1$-current and it acts on the $0$-form $1$. Although the $1$ function is not compactly supported, one can also show \cite{dR} that it makes sense to let a distribution act - through the use of a partition of unity - on a non-compactly supported test density and if the distribution itself has compact support, then this is always convergent, therefore the above expression is well-defined. If we further denote the action of a $D+1$-current $\omega$ on $1$ as
\begin{equation}
\omega[1]:=\int_M\omega,
\end{equation}we obtain "classical" notation for tensor distributions (eg. similar to what is found in \cite{Poi_Book}), since 1) it is possible to use index notation with tensor densities and make local calculations, 2) actions of distributions can be symbolically denoted by an integral. \\
\\
We identify the Heaviside step function $\theta$ with the corresponding $0$-current and define the ($1$-form) Dirac delta $\delta^\Sigma_\ast$ associated to the hypersurface $\Sigma$ to be the exterior derivative of the Heaviside current, i.e. we have for any smooth compactly supported $D$-form (or equivalently, vector density) $\varphi$
\begin{eqnarray}
\delta^\Sigma_\ast[\varphi]&=d\theta[\varphi]=-\partial\theta[\varphi]=-\theta[d\varphi]=-\int_M\theta d\varphi\nonumber \\ &=-\int_{M^+}d\varphi=-\int_{-\Sigma}\varphi=\int_\Sigma\varphi
\end{eqnarray}
Since $M$
is equipped with a volume form $\mu_{g}$, choosing a rigging $\ell^{\mu}$
pointing from $M^{-}$ to $M^{+}$ with adapted normal $n_{\mu}$
satisfying $n_{\mu}\ell^{\mu}=1$, also defines the volume element $\mu_{\ell,g}$
on $\Sigma$. Then it becomes possible to define a scalar distribution
$\delta^{\Sigma}$ evaluated on a test $D+1$-form $\varphi$ as
\begin{equation}
\delta^{\Sigma}[\varphi]:=\int_{\Sigma}f\,\mu_{\ell,g},
\end{equation}where $f$ is the scalar function uniquely determined\footnote{Note that in a shell spacetime, $\mu_g$ is merely continuous on $\Sigma$. This is not a problem as some distributions (those that can be identified with Radon measures), including the various Dirac deltas can also be seen to be linear functionals on \emph{continuous}, rather than smooth test functions \cite{dR,Tr}.} by $\varphi=f\mu_g$.
It is easy to verify the relation ($\delta^\Sigma_\mu$ are the components of the $1$-current $\delta^\Sigma_\ast$)
\begin{equation}
\delta_{\mu}^{\Sigma}=n_{\mu}\delta^{\Sigma},
\end{equation}
which shows that the scalar $\delta^{\Sigma}$ depends on the choice
of normal (i.e. it depends on the rigging $\ell$ only up to scaling).
For any soldered quantity $\bar{F}$ we then have distributionally
\begin{equation}
\partial_{\mu}\bar{F}=\overline{\partial_{\mu}F}+\left[F\right]\delta_{\mu}^{\Sigma}=\overline{\partial_{\mu}F}+\left[F\right]n_{\mu}\delta^{\Sigma}.
\end{equation}
Since the jump of the metric vanishes, the connection can be written
as
\begin{equation}
\Gamma_{\ \mu\nu}^{\kappa}=\bar{\Gamma}_{\ \mu\nu}^{\kappa},
\end{equation}
and its jump as (\ref{eq:conn_jump})
\begin{equation}
\left[\Gamma_{\ \mu\nu}^{\kappa}\right]=\frac{1}{2}\left(n_{\mu}\xi_{\nu}^{\kappa}+n_{\nu}\xi_{\mu}^{\kappa}-n^{\kappa}\xi_{\mu\nu}\right),
\end{equation}
where $\xi_{\mu\nu}=\left[g_{\mu\nu,\ell}\right]$. The curvature
tensor is then
\begin{eqnarray}
R_{\ \lambda\mu\nu}^{\kappa} & =\bar{R}_{\ \lambda\mu\nu}^{\kappa}+\delta_{\mu}^{\Sigma}\left[\Gamma_{\ \nu\lambda}^{\kappa}\right]-\delta_{\nu}^{\Sigma}\left[\Gamma_{\ \mu\lambda}^{\kappa}\right]\nonumber \\
 & =\bar{R}_{\ \lambda\mu\nu}^{\kappa}+\left(n_{\mu}\left[\Gamma_{\ \nu\lambda}^{\kappa}\right]-n_{\nu}\left[\Gamma_{\ \mu\lambda}^{\kappa}\right]\right)\delta^{\Sigma}.
\end{eqnarray}
Let $\mathcal{R}_{\ \lambda\mu\nu}^{\kappa}$ denote its singular
part, i.e. the coefficients of $\delta^{\Sigma}$. Expanding gives
\begin{equation}
\mathcal{R}_{\kappa\lambda\mu\nu}=n_{\kappa}\left[H_{\lambda\mu}\right]n_{\nu}-n_{\kappa}\left[H_{\lambda\nu}\right]n_{\mu}+n_{\lambda}\left[H_{\kappa\nu}\right]n_{\mu}-n_{\lambda}\left[H_{\kappa\mu}\right]n_{\nu},
\end{equation}
where $\left[H_{\mu\nu}\right]=\left[H_{ab}\right]\vartheta_{\mu}^{a}\vartheta_{\nu}^{b}$
(we refer to \cite{MS} for details). The scalar curvature
is calculated by contracting the curvature tensor twice as $R=\bar{R}+\mathcal{R}\delta^{\Sigma}$,
where
\begin{equation}
\mathcal{R}=2\left(\left[H_{ab}\right]\nu^{a}\nu^{b}-n^{2}\left[H\right]\right).
\end{equation}
According to the jump relations (\ref{eq:jump_rel}), we may rewrite
this as
\begin{equation}
\mathcal{R}=-2\left(\left[\chi_{ab}\right]h_{\ast}^{ab}-\left[\varphi_{a}\right]\nu^{a}\right),
\end{equation}
which is $-2\varkappa$-times the jump of the integrand of the rigged
counterterm (\ref{eq:RCT_proj}). The gravitational (scalar) Lagrangian
in the presence of a shell and interpreted as a distribution is then
\begin{equation}
L_{\mathrm{EH}}=\frac{1}{2\varkappa}R-\frac{1}{\varkappa}\left(\left[\chi_{ab}\right]h_{\ast}^{ab}-\left[\varphi_{a}\right]\nu^{a}\right)\delta^{\Sigma}.
\end{equation}
It follows that the Einstein-Hilbert action over $M$ is
\begin{eqnarray}
S_{\mathrm{EH}} & =\frac{1}{2\varkappa}\int_{M}\bar{R}\,\mu_{g}-\frac{1}{\varkappa}\int_{\Sigma}\left(\left[\chi_{ab}\right]h_{\ast}^{ab}-\left[\varphi_{a}\right]\nu^{a}\right)\mu_{\ell,g}\nonumber \\
 & =\frac{1}{2\varkappa}\int_{M^{+}}R^{+}\,\mu_{g}+\frac{1}{2\varkappa}\int_{M^{-}}R^{-}\,\mu_{g}-\frac{1}{\varkappa}\int_{\Sigma}\left[P_{\ \nu}^{\mu}\nabla_{\mu}n^{\nu}\right]\mu_{\ell,g},
\end{eqnarray}
where we have used that $\chi_{ab}h_{\ast}^{ab}-\varphi_{a}\nu^{a}$
can be written in the form $P_{\ \nu}^{\mu}\nabla_{\mu}n^{\nu}$.
If we add to this the bulk and thin shell matter actions, we obtain
the same variational principle as given by (\ref{eq:total_action}).
We have thus shown that if instead of splitting the action into separate
integrals on $M^{+}$ and $M^{-}$ and adding counterterms, we integrate
over $M$ while taking into account the singular contribution to the
Lagrangian, the resulting singular terms give precisely the difference
of the counterterms that otherwise would have had to be added by hand.\\
\\
It is interesting to note that there is no a priori reason for the
singular part of the Lagrangian to have the same value as the difference
of the counterterms. The Einstein-Hilbert Lagrangian density can be
written in the form
\begin{equation}
\mathcal{L}_{\mathrm{EH}}=P^{\kappa\lambda\mu\nu}\left(g\right)\sqrt{-\mathfrak{g}}\partial_{\kappa}\partial_{\lambda}g_{\mu\nu}+Q\left(g,\partial g\right)\sqrt{-\mathfrak{g}},
\end{equation}
where the coefficients are
\begin{equation}
P^{\kappa\lambda\mu\nu}\left(g\right)=\frac{1}{2\varkappa}\left(g^{\kappa\mu}g^{\lambda\nu}-g^{\kappa\lambda}g^{\mu\nu}\right),
\end{equation}
and
\begin{equation}
Q\left(g,\partial g\right)=\frac{1}{2\varkappa}\Gamma_{\kappa\mu\nu}\Gamma^{\kappa\mu\nu}-\frac{1}{2\varkappa}g_{\kappa\lambda}\Gamma_{\ast}^{\kappa}\Gamma_{\ast}^{\lambda},
\end{equation}
and $\Gamma_{\ast}^{\kappa}=\Gamma_{\ \mu\nu}^{\kappa}g^{\mu\nu}$.
Since only the second derivatives contribute singular terms, the Lagrangian
has the distributional form
\begin{equation}
\mathcal{L}_{\mathrm{EH}}=\bar{\mathcal{L}}_{\mathrm{EH}}+P^{\kappa\lambda\mu\nu}\sqrt{-\mathfrak{g}}n_{\kappa}n_{\lambda}\left[g_{\mu\nu,\ell}\right]\delta^{\Sigma},
\end{equation}
while a variational counterterm is given by
\begin{equation}
B=-\oint_{\partial M}n_{\kappa}n_{\lambda}P^{\kappa\lambda\mu\nu}g_{\mu\nu,\ell}\mu_{\ell,g},
\end{equation}
which is obtainable by integrating the first term in $\mathcal{L}_{\mathrm{EH}}$
by parts.\\
\\
Since $P^{\kappa\lambda\mu\nu}$ is algebraic in the metric and thus
does not depend on the transversal derivatives, it is clear that the
jump of the integrand of $B$ is the same as the singular part of
$\mathcal{L}_{\mathrm{EH}}$. However if $P^{\kappa\lambda\mu\nu}$
were to depend on the metric's transversal derivative, the singular
part of the Lagrangian would be mathematically meaningless as $P^{\kappa\lambda\mu\nu}$
would be discontinuous at $\Sigma$ where it is being evaluated. If
we choose $\left.\theta\right|_{\Sigma}=1/2$ as the value of the
step function on $\Sigma$, then the meaning of such an expression
can be salvaged as $P^{\kappa\lambda\mu\nu}$ evaluated on the average
value of the metric's transversal derivative at the price of taking
products of Dirac deltas with discontinuous functions. Moreover, were
$P^{\kappa\lambda\mu\nu}$ to depend on the metric's derivatives,
the counterterm would have to take a different form as one could no
longer get rid of second derivatives in the action by simple integrations
by parts.\\
\\
It thus seems that such a simple relation between the singular part
of the Lagrangian and the jump of the counterterms exists if the Lagrangian
is affine in the second derivatives of the field with coefficients
that do not depend on the derivatives of the field, however if these
conditions are violated in a modified gravitational theory the above
derivation breaks down and further analysis would be necessary.

\subsection{Thin shell equation from a first order action\label{subsec:Thin-shell-first}}

We mention here for completeness that the correct shell equation may
also be obtained without having to regularize the Einstein-Hilbert
action on $\Sigma$ by employing a first order equivalent. To ensure
global validity, we choose the background connection action (\ref{eq:bc_action})
rather than the noncovariant $\Gamma\Gamma$-action (\ref{eq:GG_action}).\\
\\
As we have shown in Section \ref{sec:Variational-counterterms}, we
may view the first order equivalents as the Einstein-Hilbert action
extended with a particular variational counterterm. Therefore we may
ascertain without any explicit calculations that the first order action
(\ref{eq:bc_action}) leads to the correct shell equation, as the
difference of two different variational counterterms to not depend
on the metric's transversal derivative, therefore their jumps always
agree. However it is the jump of the counterterm that appears in the
action (\ref{eq:total_action}), therefore a first order action will
lead to the same variational principle and thus the same shell equation.\\
\\
Nonetheless it is instructive to rederive the result via the first
order action from the beginning. The background connection action
(\ref{eq:bc_action}) is
\begin{equation}
S_{\nabla}=S_{\mathrm{EH}}+B_{\mathrm{BC}}=\int_{M}L_{\nabla}\,\mu_{g},
\end{equation}
where
\begin{equation}
L_{\nabla}=\frac{1}{2\varkappa}\left\{ \bar{R}+g^{\mu\nu}\left(\Delta_{\ \mu\rho}^{\kappa}\Delta_{\ \kappa\nu}^{\rho}-\Delta_{\ \kappa\rho}^{\kappa}\Delta_{\ \mu\nu}^{\rho}\right)\right\} ,
\end{equation}
and $\Delta_{\ \mu\nu}^{\kappa}=\Gamma_{\ \mu\nu}^{\kappa}-\bar{\Gamma}_{\ \mu\nu}^{\kappa}$.
To ensure manifest covariance, we consider $L_{\nabla}$ to be a function
of $g_{\mu\nu}$ and $\bar{\nabla}_{\kappa}g_{\mu\nu}$, the covariant
derivative of the metric with respect to the background connection,
where the relation \cite{Wald}
\begin{equation}
\Delta_{\ \mu\nu}^{\kappa}=\frac{1}{2}g^{\kappa\lambda}\left(\bar{\nabla}_{\mu}g_{\nu\lambda}+\bar{\nabla}_{\nu}g_{\mu\lambda}-\bar{\nabla}_{\lambda}g_{\mu\nu}\right)
\end{equation}
is relevant. A variation of the action leads symbolically to
\begin{equation}
\delta S_{\nabla}=\int_{M}\left(\delta L_{\nabla}+L_{\nabla}g^{\mu\nu}\delta g_{\mu\nu}\right)\mu_{g},
\end{equation}
where
\begin{eqnarray}
\delta L_{\nabla} & =\frac{\partial L_{\nabla}}{\partial g_{\mu\nu}}\delta g_{\mu\nu}+\frac{\partial L_{\nabla}}{\partial\bar{\nabla}_{\kappa}g_{\mu\nu}}\delta\bar{\nabla}_{\kappa}g_{\mu\nu}\nonumber \\
 & =\frac{\partial L_{\nabla}}{\partial g_{\mu\nu}}\delta g_{\mu\nu}+\frac{\partial L_{\nabla}}{\partial\bar{\nabla}_{\kappa}g_{\mu\nu}}\bar{\nabla}_{\kappa}\delta g_{\mu\nu}.
\end{eqnarray}
Since the background connection $\bar{\nabla}_{\kappa}$ has no a
priori relation with the volume element $\mu_{g}$, we may not use
Gauss' theorem with it. Therefore we must express $\bar{\nabla}_{\kappa}\delta g_{\mu\nu}$
with the Levi-Civita connection $\nabla_{\kappa}$. This is accomplished
via the difference formula \cite{Wald}
\begin{equation}
\bar{\nabla}_{\kappa}\delta g_{\mu\nu}=\nabla_{\kappa}\delta g_{\mu\nu}+\Delta_{\ \kappa\mu}^{\lambda}\delta g_{\lambda\nu}+\Delta_{\ \kappa\nu}^{\lambda}\delta g_{\mu\lambda}.
\end{equation}
Inserting this into the variation gives
\begin{equation}
\fl\delta L_{\nabla}=\left(\frac{\partial L_{\nabla}}{\partial g_{\mu\nu}}+\frac{\partial L_{\nabla}}{\partial\bar{\nabla}_{\kappa}g_{\lambda\nu}}\Delta_{\ \kappa\lambda}^{\mu}+\frac{\partial L_{\nabla}}{\partial\bar{\nabla}_{\kappa}g_{\mu\lambda}}\Delta_{\ \kappa\lambda}^{\nu}-\nabla_{\kappa}\frac{\partial L_{\nabla}}{\partial\bar{\nabla}_{\kappa}g_{\mu\nu}}\right)\delta g_{\mu\nu}+\nabla_{\kappa}\left(\frac{\partial L_{\nabla}}{\partial\bar{\nabla}_{\kappa}g_{\mu\nu}}\delta g_{\mu\nu}\right),
\end{equation}
thus the variation of the action is
\begin{eqnarray}
\delta S_{\nabla} & =\int_{M}\left(\frac{\partial L_{\nabla}}{\partial g_{\mu\nu}}+L_{\nabla}g^{\mu\nu}+\frac{\partial L_{\nabla}}{\partial\bar{\nabla}_{\kappa}g_{\lambda\nu}}\Delta_{\ \kappa\lambda}^{\mu}+\frac{\partial L_{\nabla}}{\partial\bar{\nabla}_{\kappa}g_{\mu\lambda}}\Delta_{\ \kappa\lambda}^{\nu}-\nabla_{\kappa}\frac{\partial L_{\nabla}}{\partial\bar{\nabla}_{\kappa}g_{\mu\nu}}\right)\delta g_{\mu\nu}\,\mu_{g}\nonumber \\
 & +\oint_{\partial M}n_{\kappa}\frac{\partial L_{\nabla}}{\partial\bar{\nabla}_{\kappa}g_{\mu\nu}}\delta g_{\mu\nu}\,\mu_{\ell,g}.
\end{eqnarray}
The bulk terms must be $-\frac{1}{2\varkappa}G^{\mu\nu}$ in disguise,
since the action differs from the Einstein-Hilbert action in a total
derivative term only. For the boundary term we have
\begin{equation}
2\varkappa\frac{\partial L_{\nabla}}{\partial\bar{\nabla}_{\kappa}g_{\mu\nu}}=\Delta^{\kappa\mu\nu}-\frac{1}{2}\left(g^{\lambda\nu}g^{\kappa\mu}+g^{\lambda\mu}g^{\nu\kappa}-g^{\lambda\kappa}g^{\mu\nu}\right)\Delta_{\lambda}-\frac{1}{2}g^{\mu\nu}\Delta_{\ast}^{\kappa},
\end{equation}
where $\Delta_{\lambda}=\Delta_{\ \mu\lambda}^{\mu}$ and $\Delta_{\ast}^{\kappa}=\Delta_{\ \mu\nu}^{\kappa}g^{\mu\nu}$.
The details of the analogous derivation for the $\Gamma\Gamma$-action
(\ref{eq:GG_action}) are given in Appendix 9 of \cite{Mo}
and is therefore omitted here. Let the contraction of the above expression
be denoted
\begin{equation}
\fl M^{\mu\nu}:=2\varkappa n_{\kappa}\frac{\partial L_{\nabla}}{\partial\bar{\nabla}_{\kappa}g_{\mu\nu}}=n_{\kappa}\Delta^{\kappa\mu\nu}-\frac{1}{2}\left(g^{\lambda\nu}n^{\mu}+g^{\lambda\mu}n^{\nu}-n^{\lambda}g^{\mu\nu}\right)\Delta_{\lambda}-\frac{1}{2}g^{\mu\nu}\Delta_{\ast}^{\kappa}n_{\kappa}.
\end{equation}
We rewrite the variational principle for thin shells as
%\begin{eqnarray}
%S & =\frac{1}{2\varkappa}\int_{M^{+}}\left\{ \bar{R}+g^{\mu\nu}\left(\Delta_{\ \mu\rho}^{\kappa}\Delta_{\ \kappa\nu}^{\rho}-\Delta_{\ \kappa\rho}^{\kappa}\Delta_{\ \mu\nu}^{\rho}\right)\right\} \mu_{g}+\frac{1}{2\varkappa}\int_{M^{-}}\left\{ \bar{R}+g^{\mu\nu}\left(\Delta_{\ \mu\rho}^{\kappa}\Delta_{\ \kappa\nu}^{\rho}-\Delta_{\ \kappa\rho}^{\kappa}\Delta_{\ \mu\nu}^{\rho}\right)\right\} \mu_{g}\nonumber \\
% & +\int_{M^{+}}\mathcal{L}_{\mathrm{M}}d^{D+1}x+\int_{M^{-}}\mathcal{L}_{\mathrm{M}}d^{D+1}x+\int_{\Sigma}\mathcal{L}_{\mathrm{TS}}d^{D}y,
%\end{eqnarray}
\begin{eqnarray}
\fl S =\frac{1}{2\varkappa}\int_{M^{+}}\left\{ \bar{R}+g^{\mu\nu}\left(\Delta_{\ \mu\rho}^{\kappa}\Delta_{\ \kappa\nu}^{\rho}-\Delta_{\ \kappa\rho}^{\kappa}\Delta_{\ \mu\nu}^{\rho}\right)\right\} \mu_{g} \nonumber\\ +\frac{1}{2\varkappa}\int_{M^{-}}\left\{ \bar{R}+g^{\mu\nu}\left(\Delta_{\ \mu\rho}^{\kappa}\Delta_{\ \kappa\nu}^{\rho}-\Delta_{\ \kappa\rho}^{\kappa}\Delta_{\ \mu\nu}^{\rho}\right)\right\} \mu_{g}\nonumber \\
 +\int_{M^{+}}\mathcal{L}_{\mathrm{M}}d^{D+1}x+\int_{M^{-}}\mathcal{L}_{\mathrm{M}}d^{D+1}x+\int_{\Sigma}\mathcal{L}_{\mathrm{TS}}d^{D}y,
\end{eqnarray}
and varying this with respect to the metric gives
\begin{eqnarray}
\delta S & =\int_{M^{+}}\frac{1}{2}\left(T^{\mu\nu}-\frac{1}{\varkappa}G^{\mu\nu}\right)\delta g_{\mu\nu}\,\mu_{g}+\int_{M^{-}}\frac{1}{2}\left(T^{\mu\nu}+\frac{1}{\varkappa}G^{\mu\nu}\right)\delta g_{\mu\nu}\,\mu_{g}\nonumber \\
 & -\frac{1}{2\varkappa}\int_{\Sigma}\left[M^{\mu\nu}\right]\delta g_{\mu\nu}\,\mu_{\ell,g}+\int_{\Sigma}\frac{1}{2}S^{\mu\nu}\delta g_{\mu\nu}\,\mu_{\ell,g},
\end{eqnarray}
where $S^{\mu\nu}$ is again defined by (\ref{eq:ssem_def}).
Imposing the stationarity of the action gives the boundary equation
\begin{equation}
\varkappa S^{\mu\nu}=\left[M^{\mu\nu}\right].
\end{equation}
The jump of $M^{\mu\nu}$ can be written as
\begin{equation}
\left[M^{\mu\nu}\right]=n_{\kappa}\left[\Gamma^{\kappa\mu\nu}\right]-\frac{1}{2}\left(g^{\lambda\nu}n^{\mu}+g^{\lambda\mu}n^{\nu}-n^{\lambda}g^{\mu\nu}\right)\left[\Gamma_{\lambda}\right]-\frac{1}{2}g^{\mu\nu}\left[\Gamma_{\ast}^{\kappa}\right]n_{\kappa},\label{eq:Mjump}
\end{equation}
where $\Gamma_{\lambda}=\Gamma_{\ \lambda\mu}^{\mu}$ and $\Gamma_{\ast}^{\kappa}=\Gamma_{\ \mu\nu}^{\kappa}g^{\mu\nu}$.
The connection $\bar{\nabla}_{\mu}$ was assumed to be a smooth background
structure, therefore $\left[\Delta_{\ \mu\nu}^{\kappa}\right]=\left[\Gamma_{\ \mu\nu}^{\kappa}\right]-\left[\bar{\Gamma}_{\ \mu\nu}^{\kappa}\right]=\left[\Gamma_{\ \mu\nu}^{\kappa}\right]$.
If we decompose $\left[M^{\mu\nu}\right]$ in the frame $\left(\ell,e_{a}\right)$
(since the calculation is lengthy, the details are in \ref{sec:Decomposition-of-M}),
we obtain that $\left[M^{00}\right]=0$, $\left[M^{0a}\right]=0$,
and thus $\left[M^{\mu\nu}\right]$ is tangential with $\left[M^{\mu\nu}\right]=\left[M^{ab}\right]e_{a}^{\mu}e_{b}^{\nu}$,
its projected components being
\begin{equation}
\fl\left[M^{ab}\right]=n^{2}\left(\left[H\right]h_{\ast}^{ab}-\left[H^{ab}\right]\right)+\nu^{a}\left[H_{\ c}^{b}\right]\nu^{c}+\left[H_{\ c}^{a}\right]\nu^{b}\nu^{c}-\left[H_{cd}\right]\nu^{c}\nu^{d}h_{\ast}^{ab}-\left[H\right]\nu^{a}\nu^{b},
\end{equation}
which is equal to $\left[\Pi^{ab}\right]$. The shell equation is
\begin{equation}
\varkappa S^{ab}=\left[M^{ab}\right],
\end{equation}
where $S^{ab}$ is defined by $S^{\mu\nu}=S^{ab}e_{a}^{\mu}e_{b}^{\nu}$.
This equation agrees with (\ref{eq:shelleq_final}), which shows that
the first order action indeed leads to the correct equations.\\
\\
We remark that the corresponding derivation for the Einstein-Hilbert
action extended with counterterms crucially relied on the variation
formula (\ref{eq:varformgen}) originally derived by Parattu \etal
\cite{Par}, which is a nontrivial result and difficult to obtain.
On the other hand the first order action provided a straightforward
derivation, which is clearly advantageous. The disadvantage of the
first order approach is that sufficiently complicated theories of
gravitation (eg. Horndeski theory \cite{Horn}) do not admit first
order equivalent Lagrangians, therefore this method cannot always
be relied on.\\
\\
Whether a given modified theory of gravity with second order field equations can be described in terms of a first order Lagrangian can be determined easily by looking at the field equations.  A first order Lagrangian will produce Euler-Lagrange equations that have at most an affine dependence on the second derivatives of the field variables. However it is known \cite{AD,Rossi}  that - at least locally - the converse of this statement is also true, every locally variational second-order differential equation\footnote{Strictly speaking, those differential equations for which the number of equations agree with the number of unknown functions, which are referred to as source equations in eg. \cite{Takens}. Euler-Lagrange equations are always source equations.} that is affine in the second derivatives has a local first order Lagrangian. Thus, a theory of gravitation specified in terms of a second order Lagrangian with second order field equations will have a (possibly only local and non-covariant) first order equivalent if and only if the field equations are affine functions of the second derivatives.\\
\\
Looking at the field equations of Horndeski's theory (presented for example \cite{PS}) one can ascertain that the restrictions $G_5(\phi, X)=0$, $G_4(\phi,X)=G_4(\phi)$ and $G_3(\phi,X)=G_3(\phi)$ are necessary to ensure the existence of a first order equivalent. This includes the Brans-Dicke type theories where the scalar field Lagrangian is first order and the non-minimal coupling of the scalar field to gravity does not involve the scalar field derivatives  but excludes the galileon-type models as well as kinetic gravity braiding where the higher-order nonlinear derivative interaction of the scalar field prevents the existence of first order equivalent Lagrangians. Outside Horndeski theories, Gauss-Bonnet gravity is an example of a theory with no first order Lagrangian, as the field equations  are quadratic in the curvature tensors \cite{Dav}.

\section{Conclusions}

The purpose of this paper was to provide a variational formalism for
spacetimes containing a thin shell of completely unconstrained signature.
To treat shells of arbitrary signature, we have used the formalism
of rigged hypersurfaces, reviewed in Section \ref{sec:Rigged-hypersurfaces}.
Shells are incorporated into the variational principle as interior
boundaries and their equations of motion are the natural boundary
conditions on them. The Einstein-Hilbert action needed to be regularized
at the shell to ensure a valid variational principle. We have investigated
multiple possible regularization procedures.\\
\\
In Subsection \ref{subsec:Thin-shell-ct}, regularization has been
carried out by adding variational counterterms (reviewed in Section
\ref{sec:Variational-counterterms}) to the action. The shell equation
(\ref{eq:shelleq_final}) obtained by varying this modified action
reproduces the results obtained through distributional methods by
Barrabes and Israel \cite{BI}, Mars and Senovilla \cite{MS}
and Senovilla \cite{Sen}. We have shown that the shell equation
does not depend on the choice of the counterterm and have identified
the geometric quantity the jump of which appears in the equations
of motion to be the (tensorial) canonical momentum of the gravitational
field, generalized to unconstrained instead of just spacelike foliations.\\
\\
We have considered a different regularization process in Subsection
\ref{subsec:Thin-shell-dist} by focusing on the singular part of
the Lagrangian. We have shown that the singular term is related to
the jump of the counterterm and leads to the same variational principle.
This generalized the procedure employed by eg. Hajicek and Kijowski
\cite{HK} to arbitrary shells. We have also argued that a more
general Lagrangian might have a less trivial relationship between
the singular parts and the counterterms.\\
\\
Finally, in Subsection \ref{subsec:Thin-shell-first}, we have obtained
the equations of motion of the shell by employing a first order equivalent
Lagrangian. This lead to a simpler variational procedure, but we have
noted that more complicated theories might not have first order equivalents,
rendering this method less adequate for generalization.\\
\\
Aside from filling a gap in the literature, we expect that this work
would be useful for formulating thin shells and junction conditions
along generic hypersurfaces in second order modified theories of gravity,
such as Horndeski theory \cite{Horn}. Second order Lagrangians
are capable of producing second order differential equations at least
quadratic in the second derivatives (the equations of motions associated
with the $G_{3}$ term in Horndeski's theory is an example), which
could lead to ill-defined products of delta functions if the distributional
method were to be followed. Thus it would seem that variational approaches
to thin shells are better behaved for such theories and always lead
to unambigous shell equations.

\section*{Acknowledgements}

This research was funded by the Hungarian National Research Development
and Innovation Office (NKFIH) in the form of grant 123996. I am grateful
to László Á. Gergely for valuable discussions on the topics of this
paper.

\appendix

\section*{Appendix}

\section{Decomposition of $\Pi^{\mu\nu}$}\label{sec:Decomposition-of-PI}

In this Appendix we carry out the
explicit decomposition of the tensor field
\begin{equation}
\Pi^{\mu\nu}=g^{\mu\nu}\left(P_{\ \sigma}^{\rho}\nabla_{\rho}n^{\sigma}\right)-\nabla^{(\mu}n^{\nu)}-\nabla_{\rho}\ell^{\rho}n^{\mu}n^{\nu}
\end{equation}
defined along the hypersurface $\Sigma$ in the frame $\left(\ell,e_{a}\right)$.
This calculation is best carried out by evaluating $\Pi^{\mu\nu}$
in adapted coordinates $\left(\sigma,y^{a}\right)$ such that the
$y^{a}$ parametrize $\Sigma$, while
\begin{equation}
\ell^{\mu}=\left(\frac{\partial}{\partial\sigma}\right)^{\mu}.
\end{equation}
In such an adapted coordinate system we have
\begin{eqnarray}
\ell^{\mu} & =\delta_{0}^{\mu},\quad n_{\mu}=\delta_{\mu}^{0},\nonumber \\
g^{\mu\nu} & =n^{2}\delta_{0}^{\mu}\delta_{0}^{\nu}+\nu^{a}\left(\delta_{0}^{\mu}\delta_{a}^{\nu}+\delta_{a}^{\mu}\delta_{0}^{\nu}\right)+h_{\ast}^{ab}\delta_{a}^{\mu}\delta_{b}^{\nu},\nonumber \\
n^{\mu} & =g^{\mu0}=n^{2}\delta_{0}^{\mu}+\nu^{a}\delta_{a}^{\mu},
\end{eqnarray}
\begin{equation}
P_{\ \sigma}^{\rho}\nabla_{\rho}n^{\sigma}=\chi_{cd}h_{\ast}^{cd}-\varphi_{c}\nu^{c},
\end{equation}
and the connection $\Gamma_{\ \mu\nu}^{\kappa}$ has components
\begin{equation}
\Gamma_{\ 00}^{0}=U,\quad\Gamma_{\ 00}^{a}=Z^{a},
\end{equation}
\begin{equation}
\Gamma_{\ 0a}^{0}=\varphi_{a},\quad\Gamma_{\ ab}^{0}=-\chi_{ab},\quad\Gamma_{\ 0b}^{a}=\psi_{b}^{a},\quad\Gamma_{\ ab}^{c}=\gamma_{\ ab}^{c}.
\end{equation}
The elements $U$ and $Z^{a}$ involve transversal derivatives of
the frame vectors and thus are not independent of the way the quantities
are extended off $\Sigma$. Fortunately, they will cancel. We first
evaluate what we can without fixing the free indices as
\begin{equation}
\Pi^{\mu\nu}=g^{\mu\nu}\left(\chi_{cd}h_{\ast}^{cd}-\varphi_{c}\nu^{c}\right)+g^{\mu\kappa}g^{\nu\lambda}\Gamma_{\ \kappa\lambda}^{0}-\Gamma_{\ \rho0}^{\rho}n^{\mu}n^{\nu},
\end{equation}
then we get
\begin{eqnarray}
\Pi^{00} & =n^{2}\left(\chi_{ab}h_{\ast}^{ab}-\varphi_{a}\nu^{a}\right)+\left(n^{2}\right)^{2}U+2n^{2}\varphi_{a}\nu^{a}-\chi_{ab}\nu^{a}\nu^{b}\nonumber \\
 & -\left(n^{2}\right)^{2}U-\left(n^{2}\right)^{2}\psi\nonumber \\
 & =n^{2}\left(\chi_{ab}h_{\ast}^{ab}-\varphi_{a}\nu^{a}\right)+2n^{2}\varphi_{a}\nu^{a}-\chi_{ab}\nu^{a}\nu^{b}-\left(n^{2}\right)^{2}\psi\nonumber \\
 & =\left(n^{2}h_{\ast}^{ab}-\nu^{a}\nu^{b}\right)\chi_{ab}+n^{2}\nu^{a}\varphi_{a}-\left(n^{2}\right)^{2}\psi,
\end{eqnarray}
\begin{eqnarray}
\Pi^{0a} & =\nu^{a}\left(\chi_{bc}h_{\ast}^{bc}-\varphi_{b}\nu^{b}\right)+g^{0\kappa}g^{a\lambda}\Gamma_{\ \kappa\lambda}^{0}-\Gamma_{\ \rho0}^{\rho}n^{2}\nu^{a}\nonumber \\
 & =\nu^{a}\left(\chi_{bc}h_{\ast}^{bc}-\varphi_{b}\nu^{b}\right)+n^{2}\nu^{a}U+n^{2}h_{\ast}^{ab}\varphi_{b}+\nu^{a}\nu^{b}\varphi_{b}-\nu^{b}h_{\ast}^{ac}\chi_{bc}\nonumber \\
 & -n^{2}\nu^{a}U-n^{2}\nu^{a}\psi\nonumber \\
 & =\left(\nu^{a}h_{\ast}^{bc}-\nu^{b}h_{\ast}^{ac}\right)\chi_{bc}+n^{2}h_{\ast}^{ab}\varphi_{b}-n^{2}\nu^{a}\psi,
\end{eqnarray}
\begin{eqnarray}
\Pi^{ab} & =h_{\ast}^{ab}\left(\chi_{cd}h_{\ast}^{cd}-\varphi_{c}\nu^{c}\right)+g^{a\kappa}g^{b\lambda}\Gamma_{\ \kappa\lambda}^{0}-\Gamma_{\ \rho0}^{\rho}\nu^{a}\nu^{b}\nonumber \\
 & =h_{\ast}^{ab}\left(\chi_{cd}h_{\ast}^{cd}-\varphi_{c}\nu^{c}\right)+\nu^{a}\nu^{b}U+\nu^{a}h_{\ast}^{bc}\varphi_{c}+h_{\ast}^{ac}\nu^{b}\varphi_{c}-h_{\ast}^{ac}h_{\ast}^{bd}\chi_{cd}\nonumber \\
 & -U\nu^{a}\nu^{b}-\nu^{a}\nu^{b}\psi\nonumber \\
 & =\left(h_{\ast}^{ab}h_{\ast}^{cd}-h_{\ast}^{ac}h_{\ast}^{bd}\right)\chi_{cd}+\left(\nu^{a}h_{\ast}^{bc}+h_{\ast}^{ac}\nu^{b}-h_{\ast}^{ab}\nu^{c}\right)\varphi_{c}-\nu^{a}\nu^{b}\psi.
\end{eqnarray}

\section{Decomposition of $\left[M^{\mu\nu}\right]$}\label{sec:Decomposition-of-M}

We now carry out the decomposition
in the frame $\left(\ell,e_{a}\right)$ of the tensor field
\begin{eqnarray}
\left[M^{\mu\nu}\right] & =n_{\kappa}\left[\Gamma^{\kappa\mu\nu}\right]-\frac{1}{2}\left(g^{\lambda\nu}n^{\mu}+g^{\lambda\mu}n^{\nu}-n^{\lambda}g^{\mu\nu}\right)\left[\Gamma_{\lambda}\right]-\frac{1}{2}g^{\mu\nu}\left[\Gamma_{\ast}^{\kappa}\right]n_{\kappa},
\end{eqnarray}
defined only along the hypersurface $\Sigma$, given in (\ref{eq:Mjump}).
As we have argued at (\ref{eq:dg_jump}), we may write the jump
of the metric's derivative as
\begin{equation}
\left[\partial_{\kappa}g_{\mu\nu}\right]=n_{\kappa}\xi_{\mu\nu},
\end{equation}
where $\xi_{\mu\nu}=\left[g_{\mu\nu,\ell}\right]$ is the jump of
the transversal derivative. The jump of the connection is then
\begin{equation}
\left[\Gamma_{\ \mu\nu}^{\kappa}\right]=\frac{1}{2}\left(n_{\mu}\xi_{\nu}^{\kappa}+n_{\nu}\xi_{\mu}^{\kappa}-n^{\kappa}\xi_{\mu\nu}\right).
\end{equation}
This gives
\begin{equation}
\left[\Gamma_{\lambda}\right]=\left[\Gamma_{\ \lambda\mu}^{\mu}\right]=\frac{1}{2}\left(n_{\mu}\xi_{\lambda}^{\mu}+n_{\lambda}\xi_{\mu}^{\mu}-n^{\mu}\xi_{\mu\lambda}\right)=\frac{1}{2}n_{\lambda}\xi_{\mu}^{\mu},
\end{equation}
and
\begin{equation}
\left[\Gamma_{\ast}^{\kappa}\right]=\frac{1}{2}\left(2n_{\mu}\xi^{\kappa\mu}-n^{\kappa}\xi_{\mu}^{\mu}\right)=n_{\mu}\xi^{\kappa\mu}-\frac{1}{2}n^{\kappa}\xi_{\mu}^{\mu}.
\end{equation}
We also have
\begin{equation}
n_{\kappa}\left[\Gamma_{\ \mu\nu}^{\kappa}\right]=\frac{1}{2}\left(n_{\mu}\xi_{\nu}^{\kappa}n_{\kappa}+n_{\nu}\xi_{\mu}^{\kappa}n_{\kappa}-n^{2}\xi_{\mu\nu}\right).
\end{equation}
With these, we can write
\begin{eqnarray}
\left[M^{\mu\nu}\right] & =\frac{1}{2}\left(n^{\mu}\xi^{\nu\kappa}n_{\kappa}+n^{\nu}\xi^{\mu\kappa}n_{\kappa}-n^{2}\xi^{\mu\nu}\right)-\frac{1}{4}\left(g^{\lambda\nu}n^{\mu}+g^{\lambda\mu}n^{\nu}-n^{\lambda}g^{\mu\nu}\right)n_{\lambda}\xi_{\kappa}^{\kappa} \nonumber \\ &-\frac{1}{2}g^{\mu\nu}\left(n_{\lambda}\xi^{\kappa\lambda} -\frac{1}{2}n^{\kappa}\xi_{\lambda}^{\lambda}\right)n_{\kappa}\nonumber \\
 & =\frac{1}{2}\left(n^{\mu}\xi^{\nu\kappa}n_{\kappa}+n^{\nu}\xi^{\mu\kappa}n_{\kappa}-n^{2}\xi^{\mu\nu}\right)-\frac{1}{4}\left(n^{\nu}n^{\mu}+n^{\mu}n^{\nu}-n^{2}g^{\mu\nu}\right)\xi_{\kappa}^{\kappa}\nonumber \\ &-\frac{1}{2}g^{\mu\nu}n_{\kappa}n_{\lambda}\xi^{\kappa\lambda}+\frac{1}{4}n^{2}g^{\mu\nu}\xi_{\lambda}^{\lambda}.
\end{eqnarray}
Contracting with $n_{\nu}$ gives
\begin{eqnarray}
\left[M^{\mu\nu}\right]n_{\nu} & =\frac{1}{2}\left(n^{\mu}\xi^{\nu\kappa}n_{\nu}n_{\kappa}+n^{2}\xi^{\mu\kappa}n_{\kappa}-n^{2}\xi^{\mu\nu}n_{\nu}\right)-\frac{1}{4}\left(n^{2}n^{\mu}+n^{2}n^{\mu}-n^{2}n^{\mu}\right)\xi_{\kappa}^{\kappa}\nonumber \\ &-\frac{1}{2}n^{\mu}n_{\kappa}n_{\lambda}\xi^{\kappa\lambda}+\frac{1}{4}n^{2}n^{\mu}\xi_{\lambda}^{\lambda}\nonumber \\
 & =\frac{1}{2}n^{\mu}\xi^{\nu\kappa}n_{\nu}n_{\kappa}-\frac{1}{4}n^{2}n^{\mu}\xi_{\kappa}^{\kappa}-\frac{1}{2}n^{\mu}n_{\kappa}n_{\lambda}\xi^{\kappa\lambda}+\frac{1}{4}n^{2}n^{\mu}\xi_{\lambda}^{\lambda}=0.
\end{eqnarray}
Since $\left[M^{\mu\nu}\right]$ is symmetric, this implies that it
is tangential to $\Sigma$ with $\left[M^{\mu\nu}\right]=\left[M^{ab}\right]e_{a}^{\mu}e_{b}^{\nu}$,
and these components are given by
\begin{eqnarray}
\left[M^{ab}\right] & =\vartheta_{\mu}^{a}\vartheta_{\nu}^{b}\left[M^{\mu\nu}\right]\nonumber\\
 & =\frac{1}{2}\left(\nu^{a}\xi^{\nu\kappa}\vartheta_{\nu}^{b}n_{\kappa}+\nu^{b}\xi^{\mu\kappa}\vartheta_{\mu}^{a}n_{\kappa}-n^{2}\xi^{\mu\nu}\vartheta_{\mu}^{a}\vartheta_{\nu}^{b}\right)-\frac{1}{4}\left(2\nu^{a}\nu^{b}-n^{2}h_{\ast}^{ab}\right)\xi_{\kappa}^{\kappa}\nonumber \\ &-\frac{1}{2}h_{\ast}^{ab}n_{\kappa}n_{\lambda}\xi^{\kappa\lambda}+\frac{1}{4}n^{2}h_{\ast}^{ab}\xi_{\lambda}^{\lambda}\nonumber\\
 & =\frac{1}{2}\left(\nu^{a}\vartheta_{\mu}^{b}+\nu^{b}\vartheta_{\mu}^{a}\right)\xi^{\mu\nu}n_{\nu}-\frac{1}{2}n^{2}\xi^{\mu\nu}\vartheta_{\mu}^{a}\vartheta_{\nu}^{b}+\frac{1}{2}\left(n^{2}h_{\ast}^{ab}-\nu^{a}\nu^{b}\right)\xi_{\kappa}^{\kappa}\nonumber \\ &-\frac{1}{2}h_{\ast}^{ab}n_{\kappa}n_{\lambda}\xi^{\kappa\lambda}
\end{eqnarray}
In order to proceed, we write
\begin{eqnarray}
\xi_{\mu\nu} & =\left(\vartheta_{\mu}^{a}e_{a}^{\kappa}+n_{\mu}\ell^{\kappa}\right)\left(\vartheta_{\nu}^{b}e_{b}^{\lambda}+n_{\nu}\ell^{\lambda}\right)\xi_{\kappa\lambda}\nonumber \\
 & =2\left[H_{ab}\right]\vartheta_{\mu}^{a}\vartheta_{\nu}^{b}+\xi_{a}^{\ell}\left(\vartheta_{\mu}^{a}n_{\nu}+n_{\mu}\vartheta_{\nu}^{a}\right)+\xi^{\ell}n_{\mu}n_{\nu},
\end{eqnarray}
where we have used $\left[H_{ab}\right]=\frac{1}{2}e_{a}^{\mu}e_{b}^{\nu}\xi_{\mu\nu}$
and defined
\begin{equation}
\xi_{a}^{\ell}=\xi_{\mu\nu}e_{a}^{\mu}\ell^{\nu},\quad\xi^{\ell}=\xi_{\mu\nu}\ell^{\mu}\ell^{\nu}.
\end{equation}
Then
\begin{eqnarray}
\xi^{\mu\nu}\vartheta_{\mu}^{a}n_{\nu} & =2\left[H_{b}^{a}\right]\nu^{b}+\xi_{b}^{\ell}\left(h_{\ast}^{ab}n^{2}+\nu^{a}\nu^{b}\right)+\xi^{\ell}\nu^{a}n^{2},\nonumber \\
\xi^{\mu\nu}\vartheta_{\mu}^{a}\vartheta_{\nu}^{b} & =2\left[H^{ab}\right]+\xi_{c}^{\ell}\left(h_{\ast}^{ac}\nu^{b}+\nu^{a}h_{\ast}^{bc}\right)+\xi^{\ell}\nu^{a}\nu^{b},\nonumber \\
\xi^{\mu\nu}n_{\mu}n_{\nu} & =2\left[H_{ab}\right]\nu^{a}\nu^{b}+2n^{2}\xi_{a}^{\ell}\nu^{a}+\left(n^{2}\right)^{2}\xi^{\ell},\nonumber \\
\xi_{\kappa}^{\kappa} & =2\left[H\right]+2\xi_{a}^{\ell}\nu^{a}+n^{2}\xi^{\ell},
\end{eqnarray}
and inserting these back into $\left[M^{ab}\right]$ gives
\begin{eqnarray}
\left[M^{ab}\right] & =\frac{1}{2}\left(2\nu^{a}\left[H_{c}^{b}\right]\nu^{c}+\xi_{c}^{\ell}\left(\nu^{a}h_{\ast}^{bc}n^{2}+\nu^{a}\nu^{b}\nu^{c}\right)+\xi^{\ell}\nu^{a}\nu^{b}n^{2}\right)\nonumber \\
 & +\frac{1}{2}\left(2\nu^{b}\left[H_{c}^{a}\right]\nu^{c}+\xi_{c}^{\ell}\left(\nu^{b}h_{\ast}^{ac}n^{2}+\nu^{a}\nu^{b}\nu^{c}\right)+\xi^{\ell}\nu^{a}\nu^{b}n^{2}\right)\nonumber \\
 & -\frac{1}{2}n^{2}\left(2\left[H^{ab}\right]+\xi_{c}^{\ell}\left(h_{\ast}^{ac}\nu^{b}+\nu^{a}h_{\ast}^{bc}\right)+\xi^{\ell}\nu^{a}\nu^{b}\right)\nonumber \\
 & +\frac{1}{2}\left(n^{2}h_{\ast}^{ab}-\nu^{a}\nu^{b}\right)\left(2\left[H\right]+2\xi_{c}^{\ell}\nu^{c}+n^{2}\xi^{\ell}\right)\nonumber \\
 & -\frac{1}{2}h_{\ast}^{ab}\left(2\left[H_{cd}\right]\nu^{c}\nu^{d}+2n^{2}\xi_{c}^{\ell}\nu^{c}+\left(n^{2}\right)^{2}\xi^{\ell}\right).
\end{eqnarray}
Here all terms involving $\xi_{a}^{\ell}$ and $\xi^{\ell}$ cancel,
and the remaining terms are
\begin{equation}
\fl\left[M^{ab}\right]=n^{2}\left(\left[H\right]h_{\ast}^{ab}-\left[H^{ab}\right]\right)+\nu^{a}\left[H_{c}^{b}\right]\nu^{c}+\nu^{b}\left[H_{c}^{a}\right]\nu^{c}-h_{\ast}^{ab}\left[H_{cd}\right]\nu^{c}\nu^{d}-\left[H\right]\nu^{a}\nu^{b}.
\end{equation}

\end{document}